\documentclass[11pt]{article}
\usepackage[english]{babel}
\usepackage[utf8]{inputenc}
\usepackage[T1]{fontenc}
\usepackage{indentfirst}
\usepackage{amsmath}
\usepackage{amssymb}
\usepackage{eufrak}
\usepackage{graphicx}
\usepackage{tikz}
\usepackage{psfrag}

\allowhyphens

\newcommand{\GG}{\mathcal{G}}
\newcommand{\HH}{\mathcal{H}}
\newcommand{\fii}{\varphi}
\newcommand{\valos}{\mathbb{R}}

\newcommand{\eps}{\varepsilon}

\newcommand{\ordo}{\mathcal{O}}

\newcommand{\ket}[1]{{\left|#1\right\rangle}}
\newcommand{\bra}[1]{{\left\langle #1\right|}}

\newcommand{\vev}[1]{\left\langle #1 \right\rangle}


\usepackage{ifpdf}

\ifpdf
\usepackage{epstopdf}
\usepackage[pdftex,colorlinks,urlcolor=blue,citecolor=blue,linkcolor=blue]{hyperref}
\else
\usepackage[hypertex,colorlinks,urlcolor=blue,citecolor=blue,linkcolor=blue]{hyperref}
\fi
\begin{document}
\numberwithin{equation}{section}

\title{LeClair-Mussardo series for two-point functions \\ in Integrable QFT}
\author{B. Pozsgay$^{1,2}$, I. M. Szécsényi$^{2,3}$
\\
~\\
  $^1$ Department of Theoretical Physics, Budapest University\\
	of Technology and Economics, 1111 Budapest, Budafoki \'{u}t 8, Hungary\\
$^2$ BME Statistical Field Theory Research Group, Institute of Physics,\\
	Budapest University of Technology and Economics, H-1111
        Budapest, Hungary\\
$^3$ Department of Mathematics, City, University of London, \\
         10 Northampton Square EC1V 0HB, United Kingdom 
      }

\maketitle

\abstract{
We develop a well-defined spectral representation
for two-point functions in relativistic Integrable QFT in finite density
situations, valid for space-like separations.
The resulting integral series is based on the infinite
volume, zero density form factors of the theory, and certain
statistical functions related to the distribution of Bethe roots in
the finite density background. Our final formulas are checked by
comparing them to previous partial 
results obtained in a low-temperature expansion. It is also show that in
the limit of large separations the new integral series factorizes into the product
of two LeClair-Mussardo series for one-point functions, thereby
satisfying the clustering requirement for the two-point function.
}

%----------------------------------------------------------------------------------------------------------------------
%%%%%%%%%%%%%%%%%%%%%%%%%%%%%%%%%%%%%%%%%%%%%
%----------------------------------------------------------------------------------------------------------------------

\section{Introduction}

One dimensional integrable models are special interacting many body systems, where
the eigenstates and eigenenergies can be computed with exact
methods. Various forms of the method called Bethe Ansatz have been
developed that apply to a wide variety of models including spin
chains, continuum models, and Quantum Field Theories
\cite{Korepin-book,sutherland-book,Mussardo:1992uc}. Interest in integrable models has been sparked by recent
experimental advances \cite{KiserletiOsszefogl-batchelor-foerster1}: it has become possible to measure
physical quantities of such systems both at equilibrium and in
far-from-equilibrium situations. In order to compare to experimental data it is essential to calculate
the correlation functions. 

However, computing exact correlations in Bethe
Ansatz solvable models is a notoriously difficult problem.
Depending on the models  a number of methods have been developed; they
include the Algebraic Bethe Ansatz \cite{Korepin-book}, methods based purely on
symmetry arguments (for example
the vertex
operator approach to spin chains \cite{Jimbo-Miwa-book}), or the so-called form factor
approach. Here we do not attempt to review the vast literature,
instead we focus on the form factor method.

The main idea of the form factor approach
is to evaluate two-point functions as a spectral sum over intermediate
states. By definition, the form factors are the matrix elements of local
operators on infinite volume scattering states, and they are
naturally related to finite volume on-shell matrix elements
\cite{fftcsa1,sajat-nested}.
Traditionally there are two ways to obtain the form
factors: either by solving a set of functional relations that follow
from factorized scattering and relativistic invariance 
\cite{Karowski:1978vz,Berg-Karowski-Weisz,bootstrap-osszfogl,slavnov-overlaps,Korepin-book,sajat-nested},
or by explicitly embedding the local operators into the
Algebraic Bethe Ansatz framework
\cite{PhysRevD.21.1523,PhysRevD.23.3081,maillet-inverse-scatt,goehmann-korepin-inverse}.
In many cases these methods lead to explicit and compact
representations of the form factors. 
On the other hand, the summation of the spectral series is typically a
very challenging problem, and its treatment depends on the specifics
of the physical situation.

First of all, the form factor approach was applied in massive Integrable QFT
in order to obtain two-point functions in the physical vacuum
\cite{bootstrap-osszfogl}. In these cases only the so-called
elementary form factors (matrix elements between the vacuum and a
multi-particle state) are needed, and the resulting integral series
has good convergence properties. Typically it can not be summed up
analytically, but a numerical treatment gives highly accurate results
\cite{zam_Lee_Yang}.

A completely different situation arises when there is
a finite density of excitations in the system. Examples include the ground states of
certain non-relativistic models such as the anti-ferromagnetic spin
chains or the Lieb-Liniger model, or finite temperature situations.
 Quantum quenches also
belong to this class of problems: the long-time limit of local correlation
functions can be evaluated on a finite density background given by the
so-called Generalized Gibbs Ensemble \cite{rigol-gge,essler-fagotti-quench-review,rigol-quench-review}.
In these cases the form factors and correlation functions display
different types of singular behaviour, depending on how the finite
density state is constructed. Starting with the infinite volume form
factors with a fixed number of particles one encounters the so-called kinematical
poles, whose treatment requires special care. On the other hand,
starting with a finite volume and increasing the number of particles
proportionally with the volume
can 
lead to a non-trivial scaling behaviour
for the transition matrix elements
\cite{springerlink:10.1007/BF01029221}.
In the XXZ spin chain and related models this approach was used to
compute the long distance behaviour of
correlations (see \cite{karol-hab} and references therein), by
studying the asymptotics of explicit determinant representations of
the form factors, and performing the relevant summations.

In the context of integrable QFT (iQFT) a framework was proposed in
\cite{leclair_mussardo} to deal with the kinematical poles of the form
factors in finite temperature situations.
Integral series were derived
for one-point and two-point functions. The resulting series (today
known as the LeClair-Mussardo series) are built on
the
basic form factors
with a finite number of particles. Additionally, they involve a thermodynamic function that
describes the distribution of Bethe roots, and in the case of the two-point
function a thermodynamic dressing of the energy and momenta of the
intermediate particles. 
Arguments and counter-examples against the LM series for two-point
functions were presented in 
\cite{Saleur:1999hq,Fring-LM,Doyon:finiteT1,Doyon:finiteTreview,Doyon:finiteT2},
whereas the result for the
one-point function was still believed to be true
\cite{Saleur:1999hq}.

In \cite{sajat-LM} it was shown that the LM series for one-point
functions follows from a finite volume expansion of mean values, which
uses the so-called connected limit of the infinite volume form
factors. The expansion itself was
 conjectured in full generality in \cite{fftcsa2} (see also
 \cite{Saleur:1999hq}), and for the non-relativistic Lieb-Liniger model it was proven
 using Algebraic Bethe Ansatz in  \cite{sajat-LM}.
 Finally, the expansion was proven also for iQFT in the recent work
 \cite{bajnok-diagonal}, which thus completed the proof of the LM series for
one-point functions.

On the other hand, the problem of the finite temperature or finite
density two-point functions has remained unresolved. Multiple works
performed a low-temperature expansion and obtained explicit formulas
for the first few terms
\cite{Konik:2001gf,Essler:2007jp,essler-konik-alternating-finiteT,Essler:2009zz,D22,takacs-szecsenyi-2p}. These
results were free of any singularities, and they could be understood
as the first few terms in an expansion of a hypothetical LM-type series, but it was not clear
what the general structure of this integral series should be.

As an alternative approach it was suggested in
\cite{Doyon:finiteT1,Doyon:finiteTreview,Doyon:finiteT2} that finite
temperature correlations should be computed using form factors that
take into account the dressing due to the finite density
background. This is in contrast to the logic of the LM series, which
uses the zero-density form factors calculated over the vacuum. Even
though there is clear physical motivation for this approach, the proposed
program has only been applied to free theories. Quite interestingly, a
similar picture emerged in the recent work \cite{Doyon-GHD-LM}, which
considered correlation functions in generic inhomogeneous
non-equilibrium situations within the framework of Generalized Hydrodynamics. An integral series was derived for the
large time and long distance limit of correlations, which would apply
as a special case also to static correlations in an arbitrary finite
density background. The results of \cite{Doyon-GHD-LM} only
concern the large time limit, nevertheless it is remarkable that the integral
series takes the same form as suggested by the works
\cite{Doyon:finiteT1,Doyon:finiteTreview,Doyon:finiteT2}. 

It is somewhat overlooked in the literature, that the formalism of the LeClair-Mussardo series was
already developed much earlier in the context of the infinite volume, non-relativistic
Quantum Inverse Scattering Method
\cite{Thacker-elso,PhysRevD.21.1523,creamer-thacker-wilkinson-81,creamer-thacker-wilkinson-rossz,PhysRevD.23.3081,honerkamp-LL}.
In the particular case of the 1D Bose gas explicit formulas were
obtained for the form factors of the field operators \cite{PhysRevD.21.1523}
and the particle current \cite{creamer-thacker-wilkinson-81}
from the
so-called Quantum Gelfand-Levitan method.
In \cite{creamer-thacker-wilkinson-rossz}
these were shown to satisfy a set of functional relations that are known today as ``form factor axioms'' in
iQFT. Furthermore, an integral series for the two-point function was
also derived in
\cite{PhysRevD.23.3081,creamer-thacker-wilkinson-rossz},
that has the
same structure as the LM series for one-point functions. In
these works the two-point function is treated as a composite object,
and the resulting series is built on the form factors of the bi-local
operator. In this approach there is no need for an insertion of intermediate
states, because the underlying method (the Quantum Gelfand-Levitan equation)
allows for an explicit representation of the bi-local product of field
operators in terms of the Faddeev-Zamolodchikov creation/annihilation
operators.

This program remained confined to the 1D Bose
gas, essentially due to the fact that in iQFT the form factors are
determined from the solution of the form factor axioms, their
structure is considerably more complicated (for example the operators
don't preserve particle number), and there is no efficient method to
treat the matrix elements of bi-local operators. Moreover, even in the 1D
Bose gas alternative approaches
(such as the finite volume Algebraic Bethe Ansatz)
became dominant, because they lead to intermediate formulas
that are more convenient for subsequent analytic or numerical
analysis.

In a completely independent line of research the paper
\cite{balog-tba-rederivation} developed
an alternative framework to deal with finite density states in
integrable models. Here the goal was to provide a field
theoretical derivation of the TBA equations, by computing
the mean value of the Hamiltonian density using its form factor
series. The main idea of this work is to consider smeared states
such that one does not hit the singularities of the form factors directly.
Although this work only considered the Hamiltonian density, its
derivations only rely on the general properties of the 
form factors, therefore all of the intermediate results (before
specifying the operator through its form factors) are valid for general one-point 
functions. This would imply an independent proof of the LeClair-Mussardo
series for one-point functions. We believe that this
connection has not yet been noticed in the literature. 

An important lesson of the works
\cite{PhysRevD.21.1523,creamer-thacker-wilkinson-81,creamer-thacker-wilkinson-rossz,PhysRevD.23.3081}
is that the bi-local operators satisfy the same kinematical pole equation
as the local ones. In the 1D Bose gas this was established using
explicit form factors for the bi-local operators.
Therefore, any type of regularization procedure that treats the
kinematical singularities has to work equally well for the one-point and
two-point functions.
In the present work we build on these ideas: we present arguments
for the validity of the kinematical pole equation for the bi-local operators
 even in integrable QFT, and derive a well-defined LeClair-Mussardo
 formula for the two-point function.
 
The article is composed as follows. In Section \ref{sec:2} we review 
previous approaches towards the LM series, both for the one-point and
two-point 
functions. In Section \ref{sec:LM2pt} we also formulate our main result using the form
factors of the bi-local operators. These are computed in Section \ref{sec:G}
by inserting a complete set of states between the two operators. The
properties of the bi-local form factors are studied in Section
\ref{sec:prop}. Two different compact representations for the LM
series are derived in Section \ref{sec:compact}, where we explicitly prove the
clustering property of the integral series for the two-point
function. Our results are compared in Section \ref{sec:comp} to earlier
calculations in a low-temperature expansion. Finally, Section
\ref{sec:conclusions} includes our conclusions.

\section{The LeClair-Mussardo series}

\label{sec:2}

In this work we consider one-point and two-point functions in
massive, relativistic, integrable QFT. We limit ourselves to theories with one
particle species with mass $m$. The scattering phase shift will be
denoted by $S(\theta)$, where $\theta$ is the rapidity variable.
We put forward that the generalization of our results to more particles with diagonal
scattering is straightforward, but theories with non-diagonal
scattering pose additional technical challenges, which are not
considered here.

Let us denote the incoming and outgoing scattering states as
\begin{equation}
  \begin{split}
   \ket{\theta_1,\dots,\theta_n},\qquad
    \bra{\vartheta_n,\dots,\vartheta_1},
  \end{split}
\end{equation}
where the rapidities are real numbers such that $\theta_j>\theta_k$
and $\vartheta_j>\vartheta_k$ for $j>k$.

We consider local operators $\ordo(x,t)$ in 2 dimensional Minkowski space.
Their form factors are uniquely defined as the matrix elements
\begin{equation}
  \begin{split}
     F_{n,m}(\vartheta_n,\dots,\vartheta_1|\theta_1,\dots,\theta_n)
 =\bra{\vartheta_n,\dots,\vartheta_1}\ordo\ket{\theta_1,\dots,\theta_m},
    \end{split}
  \end{equation}
  where $\ordo\equiv\ordo(0,0)$. Originally defined for sets of
  rapidities with the ordering given above, the form factor functions are
  extended analytically to the whole complex plain. 
  The analytic properties
  of these functions have been investigated in great detail in
  \cite{smirnov_ff,bootstrap-osszfogl} and they will be discussed  below.

Our goal is to derive integral representations for the mean values of
the one-point and two-point functions in finite density
situations.  A finite density state can be
characterized by a density of rapidities $\rho_{r}(\theta)$ such that
in a finite volume $L$ the number of particles between $\theta$ and
$\theta+\Delta\theta$ is $\Delta N=2\pi \rho(\theta)\Delta \theta$. As
usually we also define the density of holes $\rho_h(\theta)$, which
satisfies the integral equation
\begin{equation}
\label{alap-density}
    \rho_r(\theta)+ \rho_h(\theta)=p'(\theta)
   +\int_{-\infty}^\infty\frac{d\theta'}{2\pi} 
\varphi(\theta-\theta')\rho_r(\theta'),
\end{equation}
where $\varphi=-i \frac{d}{d\theta} \log S(\theta)$ is the scattering
kernel and $p'(\theta)=m\cosh\theta$ is the derivative of the
one-particle momentum.
We also define the filling fraction 
\begin{equation}
  \label{fdef}
f(\theta)=\frac{\rho^{(o)}(\theta)}{\rho^{(h)}(\theta)+\rho^{(o)}(\theta)}.
\end{equation}
The physical applications include the standard Gibbs and the
Generalized Gibbs (GGE) ensembles  \cite{rigol-gge}. In the first case the thermal average is
defined as
\begin{equation}
  \label{eq:Gibbs}
  \vev{\ordo_1(0,0)\ordo(x,t)_2}_{T}
  =\frac{\text{Tr}\left( e^{-\beta H} \ordo_1(0,0)\ordo_2(x,t) \right)}
 {\text{Tr}\left( e^{-\beta H} \right)},
\end{equation}
where $\beta=1/T$. Similarly, for the GGE
\begin{equation}
  \label{eq:GGE}
  \vev{\ordo_1(0,0)\ordo(x,t)_2}_{GGE}
  =\frac{\text{Tr}\left( e^{-\sum_j \beta_j Q_j} \ordo_1(0,0)\ordo_2(x,t) \right)}
 {\text{Tr}\left( e^{-\sum_j \beta_j Q_j} \right)}.
\end{equation}
In both cases the ensemble average can be simplified to a single mean
value on a representative state $\ket{\Omega}$, whose root
density $\rho_r(\theta)$  is determined by the Thermodynamic Bethe Ansatz (TBA)
equations \cite{zam-tba}. 
Defining the pseudo-energy as $\eps(\theta)=\log
\frac{\rho_h(\theta)}{\rho_r(\theta)}$, the standard TBA equation
reads
\begin{equation}
\label{TBA}
  \eps(\theta)=\beta e(\theta)-
  \int_{-\infty}^\infty\frac{d\theta'}{2\pi} 
\varphi(\theta-\theta') \log(1+e^{-\eps(\theta')}),
\end{equation}
with $e(\theta)=m\cosh\theta$ is the one-particle energy. Together
with \eqref{alap-density} this determines the Bethe root
distributions. In the case
of the GGE the eigenvalue functions of the higher charges also enter
the source terms.

Our main goal is to develop integral series for the objects 
\begin{equation}
  \bra{\Omega}\ordo_1(0)\ordo_2(x,t)\ket{\Omega}
\end{equation}
using their infinite volume, zero-density form factors, for arbitrary
root distributions. We allow for two different operators $\ordo_1$ and
$\ordo_2$, even though in practice they are typically the same or
simply just adjoints of each other.  We will restrict ourselves to space-like separations
$x^2-t^2>0$, for reasons to be discussed below.

In evaluating the two-point function there are two main difficulties that need to be solved.
First of all, the finite density state
$\ket{\Omega}$ is not a well defined object if one starts from the
infinite volume directly. Instead, different regularization schemes
need to be applied that increase the number of particles
gradually. Second, one has to deal with the singularities of the form
factors, including the disconnected pieces and the kinematical poles. Even
after the subtraction of the singular parts it is highly non-trivial
to find the remaining finite contributions.

In \ref{sec:1point} we review previous approaches to the one-point functions
that lead to the corresponding LeClair-Mussardo series. Later in
\ref{sec:2ptintro} we review previous attempts for the two-point
function and in \ref{sec:LM2pt} we formulate the main results.
However, before turning to the LM series we list here 
the main analytic properties of the form
factors of local operators,
and discuss the diagonal limit with a finite number of
particles.

A generic form factor 
can be expressed with the so-called elementary form factors by
applying the so-called crossing relation
\begin{equation}
  \begin{split}
 & F_{m,n}(\theta_{1}^{'},\dots,\theta_{m}^{'}|\theta_{1},\dots,\theta_{n})
 =F_{m-1,n+1}^{\mathcal{O}}(\theta_{1}^{'},\dots,\theta_{m-1}^{'}|\theta_{m}^{'}+i\pi,\theta_{1},\dots,\theta_{n})
 \\
 &  \qquad+\sum_{k=1}^{n}\Big(2\pi\delta(\theta_{m}^{'}-\theta_{k})\prod_{l=1}^{k-1}S(\theta_{l}-\theta_{k})
 F_{m-1,n-1}(\theta_{1}^{'},\dots,\theta_{m-1}^{'}|\theta_{1},\dots,\theta_{k-1},\theta_{k+1}\dots,\theta_{n})\Big).
 \label{eq:ffcrossing}
\end{split}
    \end{equation}
Here the second line includes the disconnected terms. The elementary
form factors satisfy the relations \cite{smirnov_ff,smirnov-ff1,kirillov-smirnov-1,kirillov-smirnov-2,kirillov-smirnov-3}

I. Lorentz transformation:
\begin{equation}
  \label{eq:shiftaxiom}
F_n(\theta_1+\Lambda,\dots,\theta_n+\Lambda)=e^{s_\ordo \Lambda}F_n(\theta_1,\dots,\theta_n),
\end{equation}
where $s_\ordo$ is the Lorentz-spin of the operator.

II. Exchange:
\begin{equation}
  F_{n}(\theta_{1},\dots,\theta_{k},\theta_{k+1},\dots,\theta_{n})=
      S(\theta_{k}-\theta_{k+1})F_{n}(\theta_{1},\dots,\theta_{k+1},\theta_{k},\dots,\theta_{n}).
      \label{eq:exchangeaxiom}
    \end{equation}

III. Cyclic permutation:
\begin{equation}
  F_{n}(\theta_{1}+2i\pi,\theta_{2},\dots,\theta_{n})
  =F_{n}(\theta_{2},\dots,\theta_{n},\theta_{1}).
  \label{eq:cyclicaxiom}
\end{equation}

IV. Kinematical singularity:
\begin{equation}
  -i\mathop{\textrm{Res}}_{\theta=\theta^{'}}F_{n+2}(\theta+i\pi,\theta^{'},\theta_{1},\dots,\theta_{n})
  =\left(1-\prod_{k=1}^{n}S(\theta-\theta_{k})\right)
  F_{n}(\theta_{1},\dots,\theta_{n}).
  \label{eq:kinematicalaxiom}
\end{equation}
There is a further relation related to the bound state structure of
the theory, but it will not be used in the present work. On the other
hand, we will assume that the form factors show the asymptotic
factorization 
property, when a subset of the rapidities is boosted to infinity:
\begin{equation}
  \label{ffcluster}
  \lim_{\Lambda\to\infty} F_n(\theta_1+\Lambda,\dots,\theta_m+\Lambda,\theta_{m+1},\dots,\theta_n)
=\frac{1}{\vev{\ordo}}F_m(\theta_1,\dots,\theta_m)F_{n-m}(\theta_{m+1},\dots,\theta_n).
\end{equation}
First observed by Smirnov \cite{smirnov_ff} and later proven in \cite{Delfino:1996nf},
 this relation holds for relevant scaling operators, and it is used to identify
solutions to the form factor axioms with concrete operators. 

For future use it is useful to display the kinematical pole relation
in the form
\begin{equation}
  \begin{split}
     & -i\mathop{\textrm{Res}}_{\vartheta_1=\theta_1}
 F_{n,m}(\vartheta_n,\dots,\vartheta_1|\theta_1,\dots,\theta_m)=\\
&\hspace{2cm}=\left(1-\prod_{k=2}^{m}S(\theta_1-\theta_{k})
\prod_{k=2}^{n}S(\vartheta_k-\vartheta_1)
\right)
  F_{n-1,m-1}(\vartheta_n,\dots,\vartheta_2|\theta_2,\dots,\theta_m).
\end{split}
 \label{eq:kinematicalaxiom1b}
\end{equation}
The diagonal limit is reached by setting $n=m$ and letting
$\vartheta_j\to\theta_j$. In this limit the form factor has an apparent $n$-fold
pole, but a straightforward calculation shows that
the residue is actually zero. It can be shown that around this point the form factor
behaves as
\begin{equation}
  \label{diagexp}
  F_{n,n}(\vartheta_n,\dots,\vartheta_2|\theta_2,\dots,\theta_n) \sim
  \frac{\sum_{i_1i_2\dots i_n}A_{i_1i_2\dots i_n}\eps_{i_1}\eps_{i_2}\dots\eps_{i_n}}{\prod_{j=1}^n \eps_j},
  \quad
  \eps_j=\vartheta_j-\theta_j,
\end{equation}
where the coefficients $A_{i_1i_2\dots i_n}$ are symmetric in the $n$
indices.

There are two natural ways to define a regularized diagonal
form factor. The so-called connected form factor is defined as 
\begin{equation}
  \label{conndef}
  F_{n,c}(\theta_1,\dots,\theta_n)\equiv
  \text{F.P. }\Big\{F_{2n}(\theta_1+\eps_1,\dots,\theta_n+\eps_n|\theta_n,\dots,\theta_1)\Big\},
\end{equation}
where F.P. stands for finite part, i.e. the terms which are free of
any singularities of the form $\eps_j/\eps_k$.
According to the expansion \eqref{diagexp}
this coincides with $n!A_{12\dots n}$. The second possibility is
to define the symmetric limit as
\begin{equation}
  \label{symmdef}
  F_{n,s}(\theta_1,\dots,\theta_n)\equiv
  \lim_{\eps\to 0}F_{2n}(\theta_1+\eps,\dots,\theta_n+\eps|\theta_n,\dots,\theta_1).
\end{equation}
Both diagonal form factors are symmetric in their variables. The linear
relations between them can be found using 
\eqref{eq:kinematicalaxiom1b}; they were studied in detail in \cite{fftcsa2}.
Further analytic properties of the diagonal form factors were studied in \cite{sajat-exactff}. 

\subsection{One-point functions}

\label{sec:1point}

In the seminal work  \cite{leclair_mussardo}  the following result was
proposed for the finite temperature one-point functions:
\begin{equation}
\label{LM}
 \vev{\ordo}_T= \sum_m \frac{1}{n!}
\int \frac{d\theta_1}{2\pi}\dots  \frac{d\theta_n}{2\pi}
\left(\prod_j \frac{1}{1+e^{\eps(\theta_j)}}\right)
F_{n,c}(\theta_1,\dots,\theta_n),
\end{equation}
where $\eps(\theta)$ is the solution of the TBA equations \eqref{TBA}
and $F_{n,c}$ are the connected diagonal form factors defined in \eqref{conndef}.

The main  idea of \cite{leclair_mussardo}  was
to consider
the finite-temperature problem in finite volume $L$ and in the low-temperature limit $T\ll
m$. In this case the volume parameter can be chosen to satisfy
$Le^{-m/T}\ll 1$ such that the partition
functions in \eqref{eq:Gibbs} are dominated by states with few
particles. In this case it is possible to perform an expansion in the
small parameter $e^{-m/T}$ such that the disconnected terms in the
numerator are canceled by the Boltzmann-sums in the denominator.
A formal calculation gives disconnected terms proportional to
$\delta(0)$, which were interpreted in \cite{leclair_mussardo} 
as diverging terms proportional to the volume.
It was argued that each order in the cluster expansion becomes finite after
the cancellation of all Dirac-deltas.
However, this procedure only kept to most divergent pieces in
$L$, and sub-leading singularities can also affect the remaining finite
answer.

A more rigorous approach was initiated in \cite{fftcsa2} which aimed
to evaluate the average  \eqref{eq:Gibbs} in the small temperature
limit by keeping all diverging pieces polynomial in $L$.
This was achieved by developing a precise description of the finite
volume diagonal matrix elements. In the following we briefly review
the results of  \cite{fftcsa2}.

Let us denote finite volume states by
\begin{equation*}
  \ket{\theta_1,\dots,\theta_N}_L.
\end{equation*}
Here it is understood that the rapidities solve the Bethe equations
\begin{equation*}
e^{iQ_j} \equiv e^{ip_jL}\prod_{k\ne j}S(\theta_j-\theta_k)=1,\qquad
  j=1\dots N,
\end{equation*}
where $p_j=p(\theta_j)=m\sinh(\theta_j)$. For transition matrix
elements it was found in \cite{fftcsa1}
\begin{equation}
   \bra{\theta'_1,\dots,\theta'_M}\ordo
   \ket{\theta_1,\dots,\theta_N}_L=
\frac{F^\ordo_{N,M}(\theta'_1,\dots,\theta'_M|\theta_1,\dots,\theta_N)}
{\sqrt{\rho_M(\theta'_1,\dots,\theta'_M)\rho_N(\theta_1,\dots,\theta_N)}}
+\ordo(e^{-\mu L}),
\label{FVFF}
\end{equation}
where $\rho_N$ and $\rho_M$ are Gaudin determinants:
\begin{equation}
  \rho_N(\theta_1,\dots,\theta_N)=\det \mathcal{J}^{ij},
\qquad\qquad
\mathcal{J}^{ij}=\frac{\partial Q_i}{\partial \theta_j}.
\label{Jdef}
\end{equation}
They can be interpreted as the density of states in rapidity space,
and in non-relativistic models as the exact norm of the Bethe Ansatz
wave function. The relation \eqref{FVFF} simply states that (apart from the physically
motivated normalization) the form factors are the same in finite and
infinite volume. For a coordinate Bethe Ansatz interpretation of this
statement see \cite{sajat-nested}.

For finite volume mean values the
following result was found in \cite{fftcsa2}:
\begin{equation}
    \label{fftcsa2-result2}
\bra{\theta_1,\dots,\theta_N}\ordo\ket{\theta_1,\dots,\theta_N}_L=
\frac{1}{\rho_N(\theta_1,\dots,\theta_N)}
\sum_{\{\theta_+\}\cup \{\theta_-\}}
F^\ordo_{2n,c}\big(\{\theta_-\}\big)
\bar\rho_{N-n}\big(\{\theta_+\}|\{\theta_-\}\big),
\end{equation}
where we used the restricted determinant
\begin{equation}
\label{restricted-density}
  \bar\rho_{N-n}(\{\theta_+\}|\{\theta_-\})=\det \mathcal{J}_+,
\end{equation}
where $\mathcal{J}_+$ is the sub-matrix of  $\mathcal{J}$ corresponding
to the particles in the set $\{\theta_+\}$. There is also an
alternative representation built on the symmetric diagonal
form factors \eqref{symmdef}, but it will not be used here.
A rigorous proof of \eqref{fftcsa2-result2} was
given for local operators of the Lieb-Liniger model in
\cite{sajat-LM},
and for relativistic QFT it was finally proven in
\cite{bajnok-diagonal}. It is important that \cite{bajnok-diagonal}
used the kinematical pole relation and the result
\eqref{FVFF} for off-diagonal matrix elements. 

In \cite{fftcsa2} the expansion \eqref{fftcsa2-result2} was used to
perform a rigorous low-temperature expansion of the Gibbs average, and
the LM series was confirmed up to third order. These calculations were performed in the regime $Le^{-m/T}\ll 1$, such
that it was enough to consider states with low particle numbers. 

An alternative, all-orders proof was
 later given in \cite{sajat-LM}. Here the idea was to consider a
representative state at some temperature (or
with some fixed root distribution) and to perform the thermodynamic limit directly on
the formula \eqref{fftcsa2-result2}. In this approach the physical
amplitudes $F^\ordo_{2n,c}\big(\{\theta_-\}\big)$ of
\eqref{fftcsa2-result2} enter the integrals
of the LM series, and the ratios of the determinants produce the
weight functions $1/(1+e^{\eps(\theta)})$. More generally it was
shown that for an arbitrary state we have
\begin{equation}
\label{LM2}
 \bra{\Omega}\ordo\ket{\Omega}= \sum_m \frac{1}{n!}
\int \frac{d\theta_1}{2\pi}\dots  \frac{d\theta_n}{2\pi}
\left(\prod_j f(\theta_j)\right)
F_{n,c}(\theta_1,\dots,\theta_n),
\end{equation}
where $f(\theta)$ is the filling fraction \eqref{fdef}. It was also
shown that the LM series can be expressed using the so-called
symmetric diagonal form factors defined in \eqref{symmdef}
as
\begin{equation}
  \label{LM3}
 \bra{\Omega}\ordo\ket{\Omega}=\sum_k \frac{1}{n!}
\int \frac{d\theta_1}{2\pi}\dots  \frac{d\theta_n}{2\pi}
\left( \prod_{j=1}^n f(\theta_j)\omega(\theta_j)\right)
F^\ordo_{2n,s}(\theta_1,\dots,\theta_n),
\end{equation}
where
\begin{equation}
  \label{omegadef}
  \omega(\theta)=\exp\left(-\int \frac{d\theta'}{2\pi} f(\theta') \varphi(\theta-\theta')\right).
\end{equation}
The two formulas
\eqref{LM2} and \eqref{LM3} can be considered to be partial
re-summations of each other. Results similar to \eqref{LM3} (including the weight function
$\omega(\theta)$) had been 
obtained for the Lieb-Liniger model using Algebraic Bethe Ansatz
\cite{korepin-LL1,Korepin-book}.

It is remarkable, that
results of the form \eqref{LM2} were obtained
much earlier in the context of the 1D Bose gas in
\cite{Thacker-elso,PhysRevD.21.1523,creamer-thacker-wilkinson-81,creamer-thacker-wilkinson-rossz,PhysRevD.23.3081}. 
These works considered both one-point and two-point functions, for
example 
the field-field correlation $\Psi^\dagger(x)\Psi(0)$, which
in the $x\to 0$ limit becomes the particle density operator. An
integral series with the same structure as the LM series 
was derived for  $\Psi^\dagger(x)\Psi(0)$, and in the $x\to 0$ limit the Yang-Yang
thermodynamics was obtained completely independently from the TBA
arguments \cite{Thacker-elso}. The main idea of these works is to introduce a
regularization involving a Galilean boost operator
\cite{Thacker-elso}, which renders the various singular terms finite.
A very important result of this approach, presented in \cite{creamer-thacker-wilkinson-rossz}, is that the
form factors of the two point function $\Psi^\dagger(x)\Psi(0)$
satisfy the same kinematical pole equation \eqref{eq:kinematicalaxiom} as those of the
local operators, see for example (3.11) of
\cite{creamer-thacker-wilkinson-rossz}.

An alternative infinite volume regularization scheme for one-point
functions was developed in
\cite{balog-tba-rederivation}, which considered smeared states to
avoid the singularities of the form factors. The main goal of
\cite{balog-tba-rederivation} was to provide a field theoretical
derivation of the TBA equations, by calculating the mean value of the
energy density operator in a finite density state.
Although the regularization in
\cite{balog-tba-rederivation}  is different from that of \cite{Thacker-elso},
the treatment of the form factors is the same and the calculation
relies only on the kinematical pole property.
The work \cite{balog-tba-rederivation} only considered the energy
current operator, but its methods could be adapted in a
straightforward way to arbitrary local operators, and this would give
an independent rigorous proof of the LM series \eqref{LM}. We believe
that this connection has not been noticed in the literature before.

\subsection{Earlier results for two-point functions in iQFT}

\label{sec:2ptintro}

In \cite{leclair_mussardo} the following series was proposed for the
finite temperature two-point functions:
\begin{equation}
  \begin{split}
&\vev{\mathcal{O}(x,t)\mathcal{O}(0,0)}_T-
\big(\vev{\mathcal{O}}_T\big)^2=\\
&\sum_{N=1}^\infty \frac{1}{N!} \sum_{\sigma_i=\pm 1}
\int \frac{d\theta_1}{2\pi}\dots \frac{d\theta_N}{2\pi}
\left[\prod_{j=1}^N f_{\sigma_j}(\theta_j) 
  \exp\Big(-\sigma_j(t\eps_j+ixk_j)\Big)\right]
\big|\bra{0}\mathcal{O}\ket{\theta_1\dots\theta_N}_{\sigma_1\dots\sigma_N}\big|^2,
\end{split}
\label{eq:2point_leclair_mussardo}
\end{equation}
where $f_{\sigma_j}(\theta_j)=1/(1+e^{-\sigma_j \eps(\theta_j)})$ and
$k_j=k(\theta_j)$, where 
$k(\theta)$ can be interpreted as the dressed momentum and it is given by  
\begin{equation}
  k(\theta)=m\sinh(\theta)+\int d\theta' \delta(\theta-\theta')
  \rho_1(\theta'),
\end{equation}
where $\rho_1(\theta)$ is the solution of the integral equation
\begin{equation}
  2\pi \rho_1(\theta)(1+e^{\eps(\theta)})
=m\cosh(\theta)+\int d\theta' \varphi(\theta-\theta') \rho_1(\theta').
\end{equation}
The form factors appearing in \eqref{eq:2point_leclair_mussardo} are
defined by
\begin{equation*}
  \bra{0}\mathcal{O}\ket{\theta_1\dots\theta_N}_{\sigma_1\dots\sigma_N}=
F^{\mathcal{O}}_N(\theta_1-i\pi\tilde{\sigma}_1,\dots,\theta_N-i\pi\tilde{\sigma}_N)\quad\quad
\tilde{\sigma}_j=(1-\sigma_j)/2 \in \{0,1\}.
\end{equation*}
It is an important feature that the $x$ and $t$ dependent phase factors
involve the dressed energies and momenta of the particles, where the
dressing is due to the finite density background. On the other hand,
the form factors are the bare quantities. 

An explicit counterexample to \eqref{eq:2point_leclair_mussardo} was
found in \cite{Saleur:1999hq}; this counter-example involved a
chemical potential, and it was not clear whether
\eqref{eq:2point_leclair_mussardo} could still hold for the $\mu=0$
case for which it was originally derived. 
In \cite{Fring-LM} the temperature dependent two-point
function of the stress-energy operator $T(x)$ was evaluated in the scaling Lee-Yang
model, in the massless limit. The results were compared to
benchmark calculations from CFT. 
The work \cite{Fring-LM} only considered
terms of the LM series up to $N=2$, but after an investigation
of the convergence properties of the series
it was concluded
that \eqref{eq:2point_leclair_mussardo}
can not be correct. In
\cite{Doyon:finiteT1,Doyon:finiteTreview,Doyon:finiteT2} it was shown
that the LM series is not correct in the Ising model for most
operators; it only works for operators with at most two free field factors.

Also, there is a central problem
 with the proposal  \eqref{eq:2point_leclair_mussardo}, which is
 independent from the previous counter-examples:
the higher terms with $N\ge 3$ involve ill-defined integrals and it is not
specified how to subtract the singular pieces. 
For $N\ge 3$ all the terms for which the $\sigma$
variables are not equal have poles in the $\theta$-variables. These poles are squared and
there is no prescription given for the subtraction of these singularities. As far as we know, this problem has not been
emphasized in the literature. The calculation of
\cite{Fring-LM} did not encounter this problem, because they only
included terms up to $N=2$, for which the pre-factor of the double pole is
zero. 

Later works attempted to perform a well-defined low-temperature
expansion of the Gibbs average for the two-point function. In
\cite{Konik:2001gf,Essler:2007jp,essler-konik-alternating-finiteT,Essler:2009zz} form factor series
were developed for spin chains and field theoretical models. The paper
\cite{Essler:2009zz} includes both finite and infinite volume
regularization, involving also a constant shift in the
rapidity parameters to treat the singularities of the form
factors; this corresponds to adding a Lorentz-boost operator into the
definition of the partition function.
Adding a boost is 
essentially the same technique that was used in 
\cite{Thacker-elso,PhysRevD.21.1523,creamer-thacker-wilkinson-81,creamer-thacker-wilkinson-rossz,PhysRevD.23.3081},
where the 
Galilean boost operator was used due to the non-relativistic
kinematics of the Lieb-Liniger model. 

The finite volume regularization was applied later in
\cite{D22,takacs-szecsenyi-2p}, where the main goal was to derive a
regularized form factor expansion irrespective of the details of the
model or the operator. The summation over Bethe states in the Gibbs
average was transformed into contour integrals such that the infinite
limit volume could be taken in a straightforward way. However, a
shortcoming of this method was that it required a term by term
analysis, and it was not evident how to express the general
higher order terms. 

All of these approaches use the bare form factors, that is, the
infinite volume zero density form factors of the theory. 
In \cite{Doyon:finiteT1,Doyon:finiteTreview,Doyon:finiteT2} it was
suggested that finite temperature correlations should be described by
dressed form factors that already take into account certain effects of
the background. New form factor axioms were set up for the finite
temperature case, and they were solved for problems involving
Majorana fermions. However, the approach has not yet been worked out
for interacting theories. On the other hand, an exact result was
computed in \cite{Doyon-GHD-LM} for the large scale correlations in
generic inhomogeneous, non-equilibrium situations, which also include
static backgrounds as special cases. Here we do not discuss the
results of \cite{Doyon-GHD-LM} in detail, as they apply to the large
time limit; however, in the Conclusions we give remarks about the
possible relations to our work.

\subsection{LeClair-Mussardo series for two-point functions}

\label{sec:LM2pt}

In the following we derive a new
form factor
series for the
two-point function in finite density situations. The structure of our
result is essentially the same as that of the formula
\eqref{LM}, therefore it can be called the LM-series for the two-point
function. The central idea is to treat the product of two local
operators as a composite object, and to investigate its matrix
elements. 

Let us define the form factors of the bi-local operator $\ordo_1(0,0)\ordo_2(x,t)$ as
\begin{equation}
  G_{n,m}^{x,t}(\vartheta_n,\dots,\vartheta_1|\theta_1,\dots,\theta_n)
  =\bra{\vartheta_n,\dots,\vartheta_1}\ordo_1(0,0)\ordo_2(x,t)\ket{\theta_1,\dots,\theta_m}.
\end{equation}
It is important that we restrict ourselves to space-like separations
$x^2-t^2>0$, in which case the two operators commute with each
other. This restriction is essential for some of our arguments to be presented
below.
The form factors are always Lorentz-invariant,
therefore we could transform any $|t|<|x|$  to $t=0$ with a
proper boost; however, the finite density state $\ket{\Omega}$ is
typically not Lorentz-invariant, therefore we reserve the possibility
of a finite time displacement.

In the following we present three independent arguments that show that
the bi-local form factors satisfy the same kinematical pole property
\eqref{eq:kinematicalaxiom1b}  as the usual form factors. 

Our first argument is based on the analytic properties of the 
Operator Product Expansion (OPE) \footnote{This argument was suggested to us by G\'abor
  Tak\'acs.} Writing the bi-local operator as the OPE
\begin{equation}
\ordo_1(0,0)\ordo_2(x,t)=\sum_j c^{\ordo_1\ordo_2}_j(x,t) \ordo_j(0,0)
\end{equation}
each term on the r.h.s. satisfies the analytic properties, therefore
their sum does too. This argument relies on the existence and the
absolute convergence of the OPE. Whereas this has not been rigorously proven,
it is generically believed to be true in QFT for space-like separations
\cite{zimmermannOPE,wilson1,wilson2,zimmermann-wilson-ope,OPE1}. 

Our second argument relies on the coordinate Bethe Ansatz wave
function. It was shown in \cite{sajat-nested} that in non-relativistic
cases the kinematical pole is easily proven by investigating the real
space integrals in the form factor. For this argument we require
$t=0$, but here we deal with only a finite number of rapidity variables
(independent of the background $\ket{\Omega}$), therefore we  can always
set $t$ to zero by an appropriate Lorentz-transformation.
We argue that the bi-local form factor satisfies the relation
\begin{equation}
  \begin{split}
 & -i\mathop{\textrm{Res}}_{\vartheta_1=\theta_1}
 G_{n,m}^{x,t}(\vartheta_n,\dots,\vartheta_1|\theta_1,\dots,\theta_m)=\\
&\hspace{2cm}=\left(1-\prod_{k=2}^{m}S(\theta_1-\theta_{k})
\prod_{k=2}^{n}S(\vartheta_k-\vartheta_1)
\right)
  G_{n-1,m-1}^{x,t}(\vartheta_n,\dots,\vartheta_2|\theta_2,\dots,\theta_m),
\end{split}
 \label{eq:kinematicalaxiom2}
\end{equation}
which has the same form as \eqref{eq:kinematicalaxiom1b}.
The singularity can be understood easily by real space calculation of
the form factors \cite{sajat-nested}. The two terms in the pole
represent divergent real space integrals when the particles with
rapidities $\theta_1$ and $\vartheta_1$ are before or behind the
operator and all the other particles, and the pre-factor reflects the
change of the phase of the wave functions as the particles with 
$\theta_1$ and $\vartheta_1$ are moved from the first position to the last.
The kinematical pole arises from 
infinite $x\to\pm \infty$ real space integrations, therefore it is not
sensitive to the precise locality properties of the operators, the only
requirement being the product should have a finite support in real
space. Similarly, relativistic effects that modify the Bethe Ansatz
wave function on small distances (comparable to the Compton wavelength of
the particles) do not play a role either, because the non-relativistic
derivation of  \cite{sajat-nested} only uses the long distance
behaviour of the wave function, which is described by the Bethe Ansatz
even in QFT. Therefore, this argument also supports the kinematical
pole property even for the bi-local operators.

The third argument is based on explicit representations of the bi-local form factors in terms
of the form factors of the individual operators;  such integral
formulas will be presented in
Section \ref{sec:G}. The kinematical pole is then evaluated explicitly
in Section \ref{sec:kin}. This derivation is mathematically rigorous,
but it leaves the physical meaning of the kinematical residue somewhat
obscure.
We believe that it is our first two arguments which provide the
physical understanding. 

\bigskip

In \ref{sec:1point} we reviewed three different approaches that lead to the
LM series for one-point functions:
\begin{itemize}
\item The infinite volume regularization of
  \cite{Thacker-elso,PhysRevD.21.1523,creamer-thacker-wilkinson-81,creamer-thacker-wilkinson-rossz,PhysRevD.23.3081}
  that uses a boost operator.
\item The infinite volume regularization of
  \cite{balog-tba-rederivation} that uses smeared states.
\item The finite volume regularization of \cite{fftcsa2} (supplied
  with the proof of \cite{bajnok-diagonal} for the diagonal matrix
  elements), and the presentation of \cite{sajat-LM} regarding the thermodynamic limit. 
  \end{itemize}
  All three approaches use only the kinematical pole property, and
  we have argued above that this property holds for the bi-local
  operators as well. 
It follows that finite density two-point functions are given by the
LeClair-Mussardo series
\begin{equation}
\label{LM2pt}
 \bra{\Omega}\ordo(0)\ordo(x,t)\ket{\Omega} = \sum_n \frac{1}{n!}
\left(\prod_j \int \frac{d\theta_j}{2\pi} f(\theta_j)\right)
G_{n,c}^{x,t}(\theta_1,\dots,\theta_n),
\end{equation}
where $G_{n,c}^{x,t}$ is the connected diagonal form
factor of the bi-local operator, which is defined as the
 finite part of
\begin{equation*}
  G_{n,n}^{x,t}(\theta_1+\eps_1,\dots,\theta_n+\eps_n|\theta_n,\dots,\theta_1),
\end{equation*}
which is free of any singularities of the form
$\eps_j/\eps_k$. Alternatively, the LM series can be expressed in the
same form as \eqref{LM3}:
\begin{equation}
\label{LM2ptB}
 \bra{\Omega}\ordo(0)\ordo(x,t)\ket{\Omega}=  \sum_n \frac{1}{n!}
\left(\prod_j \int \frac{d\theta_j}{2\pi} f(\theta_j)\omega(\theta_j)\right)
G_{n,s}^{x,t}(\theta_1,\dots,\theta_n),
\end{equation}
where
\begin{equation}
  G_{n,s}^{x,t}(\theta_1,\dots,\theta_n)=
  \lim_{\eps\to 0}  G_{n,n}^{x,t}(\theta_1+\eps,\dots,\theta_n+\eps|\theta_n,\dots,\theta_1),
\end{equation}
and $\omega(\theta)$ is given by \eqref{omegadef}. The equivalence
between \eqref{LM2pt} and \eqref{LM2ptB} is guaranteed by the theorems
for the connected and symmetric diagonal form factors presented in
\cite{fftcsa2}; these theorems only use the kinematical pole property,
therefore they apply to the two-point function as well.

Expressions \eqref{LM2pt}-\eqref{LM2ptB} are implicit: they do not specify how to compute
the form factors of the bi-local operator, they only describe how to
deal with the singularities of this object.
In the next section we also show how to compute the
bi-local form factors using those of the local operators. This is achieved by inserting a
complete set of (infinite volume) states between the two operators,
which 
 makes the LM series completely explicit. It is very important that the
composite form factors depend on the position $(x,t)$, and the
diagonal limit has to be taken by considering a full dependence on the
rapidities, including the kinematical factors involving  $(x,t)$.
Examples in Sec. \ref{sec:comp} show that this can lead to secular
terms.

We remark again, that in the non-relativistic case
an LM series of the form \eqref{LM2pt}
 has been already established
in \cite{creamer-thacker-wilkinson-rossz}, see for example eq. (3.18)
in that work, together with (3.14), which agrees with our definition
for $G_{n,c}^{x,t}$. An important difference between the two
situations is that in the 1D Bose gas the field operators
change particle number by one, and the bi-local form factors could be
obtained explicitly as sums of algebraic expressions.
On the other hand, in the 
 field theoretical case the operators can have matrix elements between
 any  $N$- and $M$-particle states, and the bi-local form factors are
obtained as an infinite integral series, to be presented in the next section.

%----------------------------------------------------------------------------------------------------------------------
%%%%%%%%%%%%%%%%%%%%%%%%%%%%%%%%%%%%%%%%%%%%%
%----------------------------------------------------------------------------------------------------------------------

\section{Form factors of the bi-local operators}

\label{sec:G}

In this Section we derive explicit integral representations for the
generic $n$-$m$ form factor
\begin{equation}
  \label{bilocal2}
  G^{x,t}_{n,m}(\vartheta_n,\dots,\vartheta_1| \theta_1,\dots,\theta_m)
  =\bra{\vartheta_n,\dots,\vartheta_1}\ordo_1(0,0)\ordo_2(x,t)\ket{\theta_1,\dots,\theta_m}
\end{equation}
by inserting a complete set of states between the two local
observables. The idea and the methods of this section are essentially
the same as in the original works by Smirnov \cite{smirnov_ff}.
The analytic properties of the resulting series are
investigated in Section \ref{sec:prop}.

A naive insertion of states would lead to an
encounter with the kinematical poles:  this happens when intermediate
rapidities approach one of the $\vartheta$ or $\theta$ variables. In
order to avoid these singularities we employ a well-defined expansion
of the local operators in terms of the Faddeev-Zamolodchikov (FZ)
creation/annihilation operators $Z^\dagger(\theta)$, $Z(\theta)$ \cite{Faddeev:FZ,zam-zam}.

The FZ operators 
satisfy the commutation relations
\begin{eqnarray}
\label{eq:ZF}
Z^{\dagger}(\theta_{1})Z^{\dagger}(\theta_{2}) & = & S(\theta_{1}-\theta_{2})Z^{\dagger}(\theta_{2})Z^{\dagger}(\theta_{1})\, ,\nonumber \\
Z(\theta_{1})Z(\theta_{2}) & = & S(\theta_{1}-\theta_{2})Z(\theta_{2})Z(\theta_{1})\, ,\nonumber \\
Z(\theta_{1})Z^{\dagger}(\theta_{2}) & = & S(\theta_{2}-\theta_{1})Z^{\dagger}(\theta_{2})Z(\theta_{1})+2\pi\delta(\theta_{1}-\theta_{2})\boldsymbol{1}\,.
\end{eqnarray}
Local operators are represented in terms of the FZ operators as
\cite{smirnov_ff,algebr-FT-operators,Gabor-initial-states-QFT-quench}
\begin{equation}
\label{kikifejt}
  \ordo(x,t)=\sum_{k,l}^\infty  H_{k,l}(x,t),
\end{equation}
where
\begin{equation}
\begin{split}
H_{k,l}(x,t)=&\frac{1}{k!l!}
\int\prod_{i=1}^{k}\frac{\mathrm{d}\theta_{i}}{2\pi}\int\prod_{j=1}^{l}\frac{\mathrm{d}\eta_{j}}{2\pi}
f_{k,l}^{\mathcal{O}}(\theta_{1},\dots,\theta_{k}|\eta_{l},\dots\eta_{1})
K_{x,t}(\{\theta\}|\{\eta\})\\
&\times Z^{\dagger}(\theta_{1})\dots Z^{\dagger}(\theta_{k})Z(\eta_{l})\dots
Z(\eta_{1}),
\end{split}
\label{eq:genff_expansion}
\end{equation}
where the functions $f$ can be expressed in terms of the form factors
\begin{equation}
f_{k,l}^{\mathcal{O}}(\theta_{1},\dots,\theta_{k}|\eta_{l},\dots,\eta_{1})
=F_{l+n}^{\mathcal{O}}(\theta_{k}+i\pi+i0,\dots,\theta_{1}+i\pi+i0,\eta_{1}-i0,\dots,\eta_{l}-i0)
\label{eq:general_ff}
\end{equation}
and $K_{x,t}$ is the phase factor due to the displacement of the
 operator:
\begin{equation}
\label{eq:Kxt}
K_{x,t}(\{\theta\}|\{\eta\})=e^{it(\sum_i E(\theta_i)-\sum_iE(\eta_i))-ix(\sum_i P(\theta_i)-\sum_iP(\eta_i))}\,,
\end{equation}
where $E(\theta)=m\cosh(\theta)$ and $P(\theta)=m\sinh(\theta)$ are
the energy end momentum of the one-particle states.  

An important ingredient in the expansion above is the presence of the
infinitesimal shifts of $\pm i0$. They  are irrelevant for the
off-diagonal form factors of $\ordo$, but they are necessary to obtain a well-defined
integral representation for the bi-local form factor \eqref{bilocal2}.

The bi-local form factor is obtained by inserting two instances of the expansion
\eqref{kikifejt}-\eqref{eq:genff_expansion} into \eqref{bilocal2} and
computing the contractions of the FZ operators using the algebra \eqref{eq:ZF}.
It is easy to see that only those terms contribute, where there are at
least $n$ FZ creation and $m$ annihilation operators. Also, there can not
be more than  $m$ annihilation (or  $n$ creation) operators immediately
acting on the ket (or bra) states. These constraints are satisfied by
the triple sum
\begin{equation}
G^{x,t}_{n,m}(\vartheta_n,\dots,\vartheta_1| \theta_1,\dots,\theta_m)=\sum_{k=0}^n\sum_{l=0}^m\sum_{p=0}^\infty \HH^{n,m}_{k,l,p} \, ,
\end{equation}
where
\begin{equation}
\HH^{n,m}_{k,l,p}=
\bra{\vartheta_n,\dots,\vartheta_1} H_{n-k,p+l}(0,0)H_{p+k,m-l}(x,t)\ket{\theta_1,\dots,\theta_m}  \, .
\end{equation}
In $\HH^{n,m}_{k,l,p}$ the indices $k$ and $l$ show the numbers of the disconnected
$\vartheta$ and $\theta$ rapidities.

We introduce the notation for the ordered set of rapidities
$\{\theta_{j_1},\theta_{j_2},\dots ,\theta_{j_n}\}=\{\theta\}_{J_<}$
and  $\{\vartheta_{j_n},\vartheta_{j_{n-1}},\dots
,\vartheta_{j_1}\}=\{\vartheta\}_{J_>}$, where the elements of $J$ are
ordered as $j_i <j_{i+1}$. $I_{n,<}$  denotes the ordered set
$\{1,2,\dots,n\}$ and $I_{n,>}$ denotes $\{n,n-1,\dots,1\}$. 

For each term in $\HH^{n,m}_{k,l,p}$ we need to evaluate a contraction of the form
\begin{equation}
  \bra{\{\vartheta\}_{I_{n,>}}}
Z^{\dagger}(\alpha_{1})\dots Z^{\dagger}(\alpha_{n-k})Z(\beta_{p+l})\dots Z(\beta_{1})
Z^{\dagger}(\mu_{1})\dots Z^{\dagger}(\mu_{p+k})Z(\nu_{m-l})\dots Z(\nu_{1})
\ket{\{\theta\}_{I_{m,<}}} \,, 
\end{equation}
where
\begin{equation}
  \begin{split}
    \ket{\{\theta\}_{I_{m,<}}}&=Z^\dagger(\theta_1)\dots
    Z^\dagger(\theta_m)\ket{0}\\
      \bra{\{\vartheta\}_{I_{m,>}}}&=\bra{0}Z(\vartheta_m)\dots Z(\vartheta_1).
  \end{split}
\end{equation}
The contractions lead to various Dirac-deltas and phase
factors, which can be treated using
straightforward calculations,
leading to
\begin{equation}
\label{eq:G_nm_final_original_contour}
\begin{split}
  G^{x,t}_{n,m}(\vartheta_{I_{n,>}}|\theta_{I_{m,<}})=&
  \sum_{p=0}^\infty  \frac{1}{p!}\prod_{i=1}^p
\int  \frac{\mathrm{d}\mu_{i}}{2\pi}
  \sum_{A^+\cup A^-=I_m}  \sum_{B^+\cup B^-=I_n} K_{x,t}(\{\mu\}_{I_p},\{\vartheta\}_{B^-}|\{\theta\}_{A^+})\\
&\times F^{\mathcal{O}_1}(\{\vartheta\}_{B^+_>}+i\pi +i0,\{\mu\}_{I_{p,<}}-i0,\{\theta\}_{A^-_<}-i0)\\
&\times F^{\mathcal{O}_2}(\{\vartheta\}_{B^-_>}+i\pi +i0,\{\mu\}_{I_{p,>}}+i\pi +i0,\{\theta\}_{A^+_<}-i0)  \\
&\times S^{<}(\{\theta\}_{A^+_<}|\{\theta\}_{A^-_<})
S^{>}(\{\vartheta\}_{B^-_>}|\{\vartheta\}_{B^+_>})
S^\leftrightarrow(\{\vartheta\}_{B^-_>}|\{\theta\}_{A^-_<}) \,,
\end{split}
\end{equation}
where
the two inner summations run over bipartite partitions of the index
sets $I_n$ and $I_m$, and we omitted the particle number subscript of
the form factors.

The phase factors $S^<$, $S^>$ and
$S^\leftrightarrow$ above result from the exchanges of the FZ operators such
that the set of particles $\{\theta_{A^-}\}$ can be connected to the operator
$\ordo_1$, and the set $\{\vartheta_{B^-}\}$ to $\ordo_2$. They are
defined as follows. For a given partitioning
$I_m=A^+\cup A^-$ we denote by
$\{\{\theta_{A^+_<}\},\{\theta_{A^-_<}\}\}$ the ordered set of
rapidities that is
obtained by concatenating the two ordered sets $\{\theta_{A^+_<}\}$
and $\{\theta_{A^-_<}\}$. The phase factor $S^<$ follows from the
rearrangement of rapidities as
\begin{equation}
\begin{split}
\ket{\{\theta\}_{I_{m,<}}}=S^{<}(\{\theta\}_{A^+_<}|\{\theta\}_{A^-_<}) \ket{\{\theta\}_{A^+_<},\{\theta\}_{A^-_<}}.
\end{split}
\label{eq:S_atrendezes_jobb}
\end{equation}
Explicitly it is given by
\begin{equation}
  S^{<}(\{\theta\}_{A^+_<}|\{\theta\}_{A^-_<})
  = \prod_{a^+_i > a^-_j} S(\theta_{a^-_j} -\theta_{a^+_i}),
  \end{equation}
where $a^+_i$, $i=1,\dots,|A^+|$ and $a^-_j$, $j=1,\dots,|A^-|$  are the indices in the sets $A^+$ and $A^-$.
Similarly, $S^>$ is defined as the phase factor coming from the rearrangement
\begin{equation}
\begin{split}
\bra{\{\vartheta\}_{I_{n,>}}}=S^{>}(\{\vartheta\}_{B^-_>}|\{\vartheta\}_{B^+_>}) \bra{\{\vartheta\}_{B^-_>},\{\vartheta\}_{B^+_>}} 
\end{split}
\label{eq:S_atrendezes_bal}
\end{equation}
with the explicit expression
\begin{equation}
  S^{>}(\{\vartheta\}_{B^-_>}|\{\vartheta\}_{B^+_>})=
  \prod_{b^+_i > b^-_j} S(\vartheta_{b^+_i} -\vartheta_{b^-_j}).
\end{equation}
Finally, the mutual phase factor $S^{\leftrightarrow}$ is given by
\begin{equation}
  S^\leftrightarrow(\{\vartheta\}_{B^-_>}|\{\theta\}_{A^-_<})=
\prod_{i\in B^-}\prod_{j\in A^-}S(\vartheta_i-\theta_j).
\end{equation}

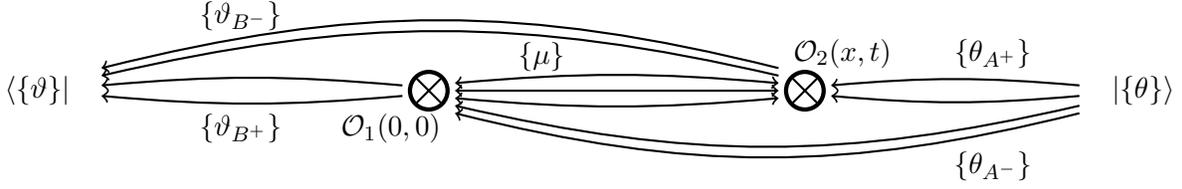
\begin{figure}
  \centering
  \begin{tikzpicture}
\newcommand{\operator}[2]{    \draw [ultra thick] (#1+0,#2+0) circle [radius=0.25];
    \draw [thick] (#1-0.17,#2-0.17) -- (#1+0.17,#2+0.17);
      \draw [thick] (#1-0.17,#2+0.17) -- (#1+0.17,#2-0.17);
    }

    \operator{0}{0}
     \operator{5}{0}
     \draw [<->,thick] (0.35,0) -- (4.65,0);
     \draw [<->,thick] (0.35,0.1) to [out=5,in=175] (4.65,0.1);
         \draw [<->,thick] (0.35,-0.1) to [out=-5,in=185]
         (4.65,-0.1);

 \draw [<-,thick] (0.35,-0.2) to [out=-13,in=193]   (8.65,-0.2);
 \draw [<-,thick] (0.35,-0.3) to [out=-14,in=194] (8.65,-0.3);

  \draw [<-,thick] (-4.35,0.2) to [out=13,in=167]   (4.65,0.2);
  \draw [<-,thick] (-4.35,0.3) to [out=14,in=166] (4.65,0.3);

  \draw [<-,thick] (5.35,0.07) to [out=5,in=175] (8.65,0.07);
  \draw [<-,thick] (5.35,-0.07) to [out=-5,in=185] (8.65,-0.07);

    \draw [<-,thick] (-4.35,0.07) to [out=5,in=175] (-0.35,0.07);
   \draw [<-,thick] (-4.35,-0.07) to [out=-5,in=185] (-0.35,-0.07);

   \node at (-0.5,-0.5) {$\ordo_1(0,0)$};
   \node at (5.5,0.5) {$\ordo_2(x,t)$};

   \node at (7.5,-1) {$\{\theta_{A^-}\}$};
   \node at (-2.5,1) {$\{\vartheta_{B^-}\}$};
   \node at (7.5,0.5) {$\{\theta_{A^+}\}$};
   \node at (-2.5,-0.5) {$\{\vartheta_{B^+}\}$};
   \node at (1.5,0.45) {$\{\mu\}$};

   \node at (9.5,0) {$\ket{\{\theta\}}$};
   \node at (-5.2,0)  {$\bra{\{\vartheta\}}$};
  \end{tikzpicture}
  \caption{Graphical interpretation of the integral representation
    \eqref{eq:G_nm_final_original_contour} for the bi-local form factor.}
  \label{fig:graphint1}
\end{figure}

The intuitive interpretation of the formula \eqref{eq:G_nm_final_original_contour} is the following: 
The incoming and outgoing particles are
connected to one of the local operators, whereas they are disconnected
from the other one; this choice is given by the partitioning of
the index sets. On the other hand, the
integration variables $\mu_j$ represent additional intermediate particles between
the two operators. The phase factors in
\eqref{eq:G_nm_final_original_contour} arise from the interchange of
positions of the particles. We stress that the $\pm i 0$ shifts for
the rapidities are essential to obtain a regular expression for the
bi-local form factor.  
A graphical representation of
\eqref{eq:G_nm_final_original_contour} is given in Fig. \ref{fig:graphint1}.

%----------------------------------------------------------------------------------------------------------------------

\bigskip

It is useful to derive an alternative representation of the bi-local
form factor, where the integration contour over the intermediate
particles is well separated from the incoming and outgoing
rapidities.
To this order we shift the integration contour in
\eqref{eq:G_nm_final_original_contour}  to $\valos
+i \alpha$ with $\alpha$ being a small real number. The advantage of
this step is that afterwards the kinematical poles at $\vartheta_j\to\theta_k$
can be studied directly, without paying attention to the
$\mu$-integrals. On the other hand, if some of the $\mu$-integrals
would run between the rapidities $\vartheta_j$ and $\theta_k$, then
the contour could be pinched, and such cases would complicate the
analysis of the kinematical singularities.

First we assume that
$\alpha>0$, then the contour shift
picks up poles of the form factors of $\ordo_1$. After a
straightforward, but tedious calculation we obtain the formula
\begin{equation}
\label{eq:G_nm_final_contour+}
\begin{split}
G^{x,t}_{n,m}(\vartheta_{I_{n,>}}|\theta_{I_{m,<}})=&
\sum_{p=0}^\infty
G^{x,t}_{n,m,p}(\vartheta_{I_{n,>}}|\theta_{I_{m,<}}), \quad \text{where}\\
G^{x,t}_{n,m,p}(\vartheta_{I_{n,>}}|\theta_{I_{m,<}})=&
\frac{1}{p!}
\prod_{i=1}^p \int_{\valos+i\alpha}\frac{\mathrm{d}\mu_{i}}{2\pi}   \sum_{A^+\cup A^-=I_m}  \sum_{B^+\cup B^-=I_n} K_{x,t}(\{\mu\}_{I_p},\{\vartheta\}_{B^-}|\{\theta\}_{A^+})\\
&\times F^{\mathcal{O}_1}(\{\vartheta\}_{B^+_>}+i\pi,\{\mu\}_{I_{p,<}},\{\theta\}_{A^-_<})
 F^{\mathcal{O}_2}(\{\mu\}_{I_{p,>}}+i\pi,\{\vartheta\}_{B^-_>}+i\pi ,\{\theta\}_{A^+_<})  \\
&\times S^{<}(\{\theta\}_{A^+_<}|\{\theta\}_{A^-_<}) S^{>}(\{\vartheta\}_{B^+_>}|\{\vartheta\}_{B^-_>})  \,,
\end{split}
\end{equation}
where
we omitted the $\pm i0$ shifts, because they are not relevant 
anymore.
Details of this calculation are presented in Appendix
\ref{appA}; here we just note that the changes in the phase factors 
(including the reordering of the rapidities in the form factor
$F^{\ordo_2}$) at a given $p$ result from adding pole contributions
from terms with $p'>p$.

We can also shift the contours to the negative
imaginary direction, which results in the alternative formula
\begin{equation}
\label{eq:G_nm_final_contour-}
\begin{split}
  G^{x,t}_{n,m,p}(\vartheta_{I_{n,>}}|\theta_{I_{m,<}})=&
  \frac{1}{p!}\prod_{i=1}^p \int_{\valos-i\alpha}\frac{\mathrm{d}\mu_{i}}{2\pi}
  \sum_{A^+\cup A^-=I_m}  \sum_{B^+\cup B^-=I_n} K_{x,t}(\{\mu\}_{I_p},\{\vartheta\}_{B^-}|\{\theta\}_{A^+})\\
&\times F^{\mathcal{O}_1}(\{\vartheta\}_{B^+_>}+i\pi,\{\theta\}_{A^-_<},\{\mu\}_{I_{p,<}})
 F^{\mathcal{O}_2}(\{\vartheta\}_{B^-_>}+i\pi ,\{\mu\}_{I_{p,>}}+i\pi,\{\theta\}_{A^+_<})  \\
&\times S^{<}(\{\theta\}_{A^-_<}|\{\theta\}_{A^+_<}) S^{>}(\{\vartheta\}_{B^-_>}|\{\vartheta\}_{B^+_>})  \,.
\end{split}
\end{equation}
The differences between the
integrands in \eqref{eq:G_nm_final_original_contour}
and those of \eqref{eq:G_nm_final_contour+} and
\eqref{eq:G_nm_final_contour-} lie in the phase factors, including the 
ordering of the rapidities in the form factors functions.

The convergence of the $\mu$-integrals in the above expressions
depends on the separation between the operators. The $\mu$-dependence
for large $|\mu|$ is given solely by the $K$-factors, given that the
asymptotic factorization property \eqref{ffcluster} holds.
It follows that the integrations
 towards $\mu\to\pm\infty$ are convergent in
\eqref{eq:G_nm_final_contour+}
if  $x<0$, and in \eqref{eq:G_nm_final_contour-} if
$x>0$. If the two operators $\ordo_{1,2}$ are not identical then the
two-point function might not have the $x\to -x$ symmetry, and only one
of the representations can be used, depending on the sign of $x$.
%In
%principle the $\mu$-integrals could be pulled back to the real 
%axis for $|\mu|>\Lambda$, where $\Lambda$ is bigger than all the
%$\theta_j$ and $\vartheta_j$ variables, or the infinite integration could be assigned a
%finite value based on this pull-back. However, this would just
%complicate the technical details, and we simply require to choose the
%appropriate contour corresponding to the sign of $x$.
On the other hand, the $\mu$-integrals would never be convergent for
time-like separations: they would diverge either for $\mu\to+\infty$ or
$\mu\to -\infty$.

It is important that the integral series is effectively a large-distance expansion
for the bi-local form factor: each $\mu$-integral carries a weight of
$e^{-ms}$ with $s^2=x^2-t^2$. This is analogous to the case of the
vacuum two-point function.

We note that essentially the same integral representations as presented in this
section were 
used by Smirnov to prove the local commutativity theorem
\cite{smirnov_ff}. This theorem states that if there are two operators
such that the form factors satisfy the axioms as given above, then the
two operators commute for space-like separations.
We do not replicate the proof here. Instead, we just point that it also uses
a shift of the integration contours in the complex
plain, which is only possible for space-like displacement between the operators.

%----------------------------------------------------------------------------------------------------------------------

\section{Form factor properties of the bi-local form factors}

\label{sec:prop}

In this Section we investigate the analytic properties of the bi-local
form factors based on the integral
representations
\eqref{eq:G_nm_final_contour+}-\eqref{eq:G_nm_final_contour-}.
Most importantly we confirm explicitly that the bi-local form factor
satisfies the kinematical pole relation \eqref{eq:kinematicalaxiom2}, which is the basis
for the derivation of the LeClair-Mussardo series. In
\ref{sec:ffdiscussion} we also discuss the implications of the form
factor axioms for the bi-local operators.

%----------------------------------------------------------------------------------------------------------------------

\subsection{Lorentz transformation}

After a Lorentz transformation every rapidity is shifted by
$\Lambda$. The S-matrix factors are invariant under the shift, since
they only depend on rapidity differences. Each elementary form factor
satisfies the Lorentz transformation axiom 
\eqref{eq:shiftaxiom}.
The phase factor transforms as 
\begin{equation}
K_{x,t}(\{\eta\}+\Lambda |\{\zeta\}+\Lambda)=K_{x',t'}(\{\eta\} |\{\zeta\})\,,
\end{equation}
where $x'=x\cosh(\Lambda)-t\sinh(\Lambda)$ and $t'=t\cosh(\Lambda)-x\sinh(\Lambda)$ are exactly the transformation rules for space-time coordinates under Lorentz transformation. As a consequence, the bi-local form factor satisfies the following Lorentz transformation rule
\begin{equation}
G^{x,t}_{n,m}(\{\vartheta\}_{I_{n,>}}+\Lambda|\{\theta\}_{I_{m,<}}+\Lambda) =e^{(s_{\ordo_1}+s_{\ordo_2})\Lambda}G^{x',t'}_{n,m}(\{\vartheta\}_{I_{n,>}}|\{\theta\}_{I_{m,<}})\,,
\end{equation}
where $s_{\ordo_i}$ is the Lorentz spin of the operator $\ordo_i$.

%----------------------------------------------------------------------------------------------------------------------

\subsection{Exchange property} 

It is easy to see that the form factors of the bi-local operators satisfy the exchange property for rapidities both sides of the operators
\begin{equation}
\begin{split}
G^{x,t}_{n,m}(\dots,\vartheta_{i+1},\vartheta_{i},\dots|\dots)=&S(\vartheta_{i+1}-\vartheta_{i}) G^{x,t}_{n,m}(\dots,\vartheta_{i},\vartheta_{i+1}|\dots) \, , \\
G^{x,t}_{n,m}(\dots|\dots,\theta_{i},\theta_{i+1},\dots)=&S(\theta_{i}-\theta_{i+1}) G^{x,t}_{n,m}(\dots|\dots,\theta_{i+1},\theta_{i},\dots) \, .
\end{split}
\end{equation}
For those terms where the rapidities to be exchanged are
in the same sets $A^\pm$ or $B^\pm$  the property follows from the property of the
elementary form factors \eqref{eq:exchangeaxiom}. In those cases,
where the rapidities are in different sets, the extra $S$-matrix
factor exactly cancels out or introduces the terms in the
$S^<({\theta}_{A^+_<}|{\theta}_{A^-_<})$ and
$S^>({\vartheta}_{B^+_>}|{\vartheta}_{B^-_>})$ phase factors to get
the appropriate ordering.

%----------------------------------------------------------------------------------------------------------------------

\subsection{Kinematical poles}

\label{sec:kin}

The kinematical pole property \eqref{eq:kinematicalaxiom} is the
essential ingredient for the proof of the LeClair-Mussardo
formula. Here we prove that it holds for the form factors of the
bi-local operators too. We use the representation
\eqref{eq:G_nm_final_contour+} in which
there are no singularities associated with the
intermediate particles. However, the
elementary form factors in the summation
\eqref{eq:G_nm_final_contour+}  have direct kinematical poles for
$\vartheta_1\sim\theta_1$, and these need to be summed up to obtain
the total residue of the bi-local form factor. 

The singularities come from terms where $\vartheta_1$ and $\theta_1$
are in the same elementary form factors, namely if
$\vartheta_1\in\{\vartheta\}_{B^+}$ and $\theta_1\in\{\theta\}_{A^-}$
or $\vartheta_1\in\{\vartheta\}_{B^-}$ and
$\theta_1\in\{\theta\}_{A^+}$. Introducing the notation
$\tilde{A}^\pm=A^\pm\setminus \{1\}$ and $\tilde{B}^\pm=B^\pm\setminus
\{1\}$, the residues in the first situation are 
\begin{equation}
\begin{split}
&\sum_{p=0}^\infty  \frac{1}{p!}\int_{\valos+i\alpha}\prod_{i=1}^p \frac{\mathrm{d}\mu_{i}}{2\pi}   \sum_{\tilde{A}^+\cup \tilde{A}^-=I_m\setminus\{1\}}  \sum_{\tilde{B}^+\cup \tilde{B}^-=I_n\setminus\{1\}} K_{x,t}(\{\mu\}_{I_p},\{\vartheta\}_{\tilde{B}^-}|\{\theta\}_{\tilde{A}^+})\\
&\times
F^{\mathcal{O}_1}(\{\vartheta\}_{\tilde{B}^+_>}+i\pi,\{\mu\}_{I_{p,<}},\{\theta\}_{\tilde{A}^-_<})\\
&\times i[S^\leftrightarrow (\{\mu\}_{I_p} |\theta_1)-S^\leftrightarrow (\theta_1 |\{\vartheta\}_{\tilde{B}^+}+i\pi) S^\leftrightarrow (\theta_1 |\{\theta\}_{\tilde{A}^-})] \\
&\times F^{\mathcal{O}_2}(\{\mu\}_{I_{p,>}}+i\pi,\{\vartheta\}_{\tilde{B}^-_>}+i\pi ,\{\theta\}_{\tilde{A}^+_<})  \\
&\times S^{<}(\{\theta\}_{\tilde{A}^+_<}|\{\theta\}_{\tilde{A}^-_<})
S^{>}(\{\vartheta\}_{\tilde{B}^+_>}|\{\vartheta\}_{\tilde{B}^-_>})
 S^\leftrightarrow (\theta_1 |\{\theta\}_{\tilde{A}^+}) S^\leftrightarrow (\{\vartheta\}_{\tilde{B}^-}|\theta_1) \,,
\end{split}
\end{equation}
while in the second situation they take the form
\begin{equation}
\begin{split}
&\sum_{p=0}^\infty  \frac{1}{p!}\int_{\valos+i\alpha}\prod_{i=1}^p \frac{\mathrm{d}\mu_{i}}{2\pi}   \sum_{\tilde{A}^+\cup \tilde{A}^-=I_m\setminus\{1\}}  \sum_{\tilde{B}^+\cup \tilde{B}^-=I_n\setminus\{1\}} K_{x,t}(\{\mu\}_{I_p},\{\vartheta\}_{\tilde{B}^-}|\{\theta\}_{\tilde{A}^+})\\
&\times
F^{\mathcal{O}_1}(\{\vartheta\}_{\tilde{B}^+_>}+i\pi,\{\mu\}_{I_{p,<}},\{\theta\}_{\tilde{A}^-_<})
F^{\mathcal{O}_2}(\{\mu\}_{I_{p,>}}+i\pi,\{\vartheta\}_{\tilde{B}^-_>}+i\pi ,\{\theta\}_{\tilde{A}^+_<})  \\
&\times i[1-S^\leftrightarrow (\theta_1 |\{\mu\}_{I_p}+i\pi) S^\leftrightarrow (\theta_1 |\{\vartheta\}_{\tilde{B}^-}+i\pi)  S^\leftrightarrow (\theta_1 |\{\theta\}_{\tilde{A}^+})] \\
&\times S^{<}(\{\theta\}_{\tilde{A}^+_<}|\{\theta\}_{\tilde{A}^-_<}) S^{>}(\{\vartheta\}_{\tilde{B}^+_>}|\{\vartheta\}_{\tilde{B}^-_>})  \,.
\end{split}
\end{equation}

Adding up the two kinds of residues we arrive to the kinematical pole property of the bi-local form factor
\begin{equation}
\begin{split}
  \mathop{\textrm{Res}}_{\vartheta_1=\theta_1}G^{x,t}_{n,m}(\{\vartheta\}_{(I_{n}\setminus\{1\})_>},\vartheta_1|\theta_1,\{\theta\}_{(I_{m}\setminus\{1\}))_<})&=
  i\left(1-
    S^\leftrightarrow(\theta_1|\{\theta\}_{I_{m}\setminus\{1\}})
    S^\leftrightarrow(\{\vartheta\}_{I_{n}\setminus\{1\}}|\theta_1)\right) \\
&\times G^{x,t}_{n,m}(\{\vartheta\}_{(I_{n}\setminus\{1\})_>}|\{\theta\}_{(I_{m}\setminus\{1\})_<}) \,.
\end{split}
\label{eq:kinematical_pole_bilocal}
\end{equation}
This is identical to the crossed version \eqref{eq:kinematicalaxiom2} of the form factor kinematical pole axiom
\eqref{eq:kinematicalaxiom}.

We remark that an analogous calculation can be performed also for the
alternative representation
\eqref{eq:G_nm_final_contour-}.

\subsection{Other properties and discussion}

\label{sec:ffdiscussion}

We have checked that the crossing and periodicity properties \eqref{eq:ffcrossing} and
\eqref{eq:cyclicaxiom}
also hold for 
the bi-local form factors. However, these relations are not relevant
for the LM series, therefore we refrain from presenting the proof. 
We also remark that the asymptotic factorization
property \eqref{ffcluster} does not hold in the bi-local case, which
can be understood simply from a physical point of view, or from the
integral representations.

So far we have argued that the bi-local form factors satisfy all
axioms that were originally derived for the local ones. Here we discuss the
implications of this statement. 

First of all, it is an interesting idea to apply the local commutativity theorem
to the composite objects. This theorem by Smirnov states that if the
form factors of two operators satisfy the set of axioms presented in
Section \ref{sec:2}, then they commute at space-like separations \cite{smirnov_ff}. On
the other hand,
our results suggest an extension of this theorem to bi-local objects. 
Such a theorem would state
that
for any three operators with the required form factor properties we have
\begin{equation}
  \label{comm3}
  [\ordo_1(x_1,t_1),\ordo_2(x_2,t_2)\ordo_3(x_3,t_3)]=0,
\end{equation}
given that all three separations are space-like. The trivial way of proving
\eqref{comm3} is to apply Smirnov's theorem twice to commute the two objects.
However, a second proof can be given by using the form factor properties of the
bi-local operator, and then repeating Smirnov's calculation for the
composite object. We refrain from presenting a rigorous proof of this
extended theorem, as it is not relevant to the present
work. Nevertheless,  we stress that the local commutativity theorem
can not be applied backwards: it does not imply that only the local
operators can satisfy the form factor axioms. 

Finally we comment on the possibility of a form factor bootstrap for
the bi-local objects. In the case of the local operators the form
factor axioms (together with the asymptotic factorization property) include enough
constraints to fix the form factors completely, both in relativistic
and non-relativistic situations. The reason for this is the following:
The form factors can always be written as a product of a 
``minimal'' part (which is two-particle irreducible) and a physical amplitude, which is a symmetric polynomial
in the appropriate variables. Then the kinematical pole axiom is used
to fix this polynomial. 
In the bi-local case the polynomial would depend also on the
displacement between the two operators, and the same number of
constraints can not fix it. Therefore,
the standard bootstrap procedure can not be applied to the bi-local form factors.

%----------------------------------------------------------------------------------------------------------------------
%%%%%%%%%%%%%%%%%%%%%%%%%%%%%%%%%%%%%%%%%%%%%
%----------------------------------------------------------------------------------------------------------------------

\section{Compact representations of the LeClair-Mussardo series}

\label{sec:compact}

Formulas \eqref{eq:G_nm_final_contour+}-\eqref{eq:G_nm_final_contour-}
give an integral representation for $G^{x,t}_{n,m}$
 in terms of the elementary form factors. The LM series is then given
 by \eqref{LM2pt} and \eqref{LM2ptB}, and this gives a well-defined
 way to
 evaluate the two-point function.

The final integral formulas \eqref{LM2pt} and \eqref{LM2ptB} require
the evaluation of the connected and symmetric diagonal form
factors. It follows from the proof of the kinematical pole property, that these operations can be
performed on the objects $G^{x,t}_{n,n,p}$ for each $n$ and $p$ separately. On the other
hand,
 computing the diagonal limit term by term in
 \eqref{eq:G_nm_final_contour+}-\eqref{eq:G_nm_final_contour-} leads 
 to singularities, 
and it is only the sum over the partitions in 
\eqref{eq:G_nm_final_contour+}-\eqref{eq:G_nm_final_contour-}
 that has the
 desired residue structure
 \begin{equation}
   \label{Gdiagexp}
  G^{x,t}_{n,n}(\vartheta_{I_{n,>}}|\theta_{I_{n,<}}) \sim
  \frac{\sum_{i_1i_2\dots i_n}A_{i_1i_2\dots i_n}\eps_{i_1}\eps_{i_2}\dots\eps_{i_n}}{\prod_{j=1}^n \eps_j},
  \quad
  \eps_j=\vartheta_j-\theta_j.
\end{equation}

One way towards singularity free expressions is to develop an integral
formula which automatically produces the diagonal 
limit of the form factors. This can be achieved by introducing
auxiliary integration variables.
 It follows from \eqref{Gdiagexp} that
\begin{equation}
  G^{x,t}_{n,c}(\theta)=
  \prod_{j=1}^n \int_{C_j} \frac{d\vartheta_j}{2\pi i}
\frac{G^{x,t}_{n,n}(\vartheta_{I_{n,>}}|\theta_{I_{n,<}})}{\prod_{k=1}^n \sinh(\vartheta_k-\theta_k)},
\end{equation}
where $C_j$ are small contours around $\theta_j$. These contours can
be opened to encircle the whole real line. Opening the contours we do not
encounter additional $n$-fold poles, therefore we can write
\begin{equation}
  G^{x,t}_{n,c}(\theta)=
\frac{1}{n!}  \prod_{j=1}^n \int_{C} \frac{d\vartheta_j}{2\pi i}
\frac{G^{x,t}_{n,n}(\vartheta_{I_{n,>}}|\theta_{I_{n,<}})}{\prod_{k=1}^n \sinh(\vartheta_k-\theta_k)},
\end{equation}
where $C$ is a narrow contour around the real line and the factor of $1/n!$ has been inserted due to the possible
permutations of the set $\{\vartheta\}$.
It is important that
$C$ is narrow enough so that it does not hit the integration
contours $\valos+i\alpha$ or $\valos-i\alpha$ for the intermediate
particles in \eqref{eq:G_nm_final_contour+}-\eqref{eq:G_nm_final_contour-}.

It follows that the LeClair-Mussardo series can be written finally as
\begin{equation}
\label{LMmaskepp}
  \begin{split}
&    \bra{\Omega}\ordo_1(0)\ordo_2(x,t)\ket{\Omega}=\sum_{n=0}^\infty
\frac{1}{(n!)^2} \prod_{j=1}^n \left[
  \int_C \frac{d\vartheta_j}{2\pi i} \int_\valos \frac{d\theta_j}{2\pi}f(\theta_j)
\right]
\frac{G^{x,t}_{n,n}(\vartheta_{I_{n,>}}|\theta_{I_{n,<}})}{\prod_{k=1}^n \sinh(\vartheta_k-\theta_k)}.
  \end{split}
\end{equation}

An alternative compact formula can be given by using the symmetric
diagonal limit in \eqref{LM2ptB}. The idea is to exchange the $\eps\to
0$ limit with the summations. This is justified (at least in the
thermal situation) due to the
exponentially decaying factors $e^{-mR\cosh(\theta)}$ and
$e^{-mx\cosh(\theta)}$ associated with the numbers $n$ and $p$, and
the limits on the growth of the form factors following from the
asymptotic factorization property \eqref{ffcluster}. We then
obtain
\begin{equation}
  \label{LM2ptC}
  \begin{split}
 & \bra{\Omega}\ordo_1(0)\ordo_2(x,t)\ket{\Omega}=\\
  &=\lim_{\eps\to 0}  \left[\sum_n \frac{1}{n!}
\left(\prod_j \int \frac{d\theta_j}{2\pi} f(\theta_j)\omega(\theta_j)\right)
G_{n,n}^{x,t}(\theta_1+\eps,\dots,\theta_n+\eps|\theta_n,\dots,\theta_1)\right].
\end{split}
\end{equation}
Substituting for example  \eqref{eq:G_nm_final_contour+} gives
\begin{equation}
  \label{LM2ptD}
  \begin{split}
 & \bra{\Omega}\ordo_1(0)\ordo_2(x,t)\ket{\Omega}=\lim_{\eps\to 0}
 \left[\sum_{n,p=0}^\infty \frac{1}{n!} \frac{1}{p!}
\prod_{j=1}^n \int_\valos \frac{d\theta_j}{2\pi} f(\theta_j)\omega(\theta_j)
\prod_{i=1}^p \int_{\valos+i\alpha}\frac{\mathrm{d}\mu_{i}}{2\pi} \right.\times \\
&\times \sum_{A^+\cup A^-=I_n}  \sum_{B^+\cup B^-=I_n} K_{x,t}(\{\mu\}_{I_p},\{\theta\}_{B^-}+i\eps|\{\theta\}_{A^+})
 S^{<}(\{\theta\}_{A^+_<}|\{\theta\}_{A^-_<})
S^{>}(\{\theta\}_{B^+_>}|\{\theta\}_{B^-_>})\\
&\times \left.F^{\ordo_1}(\{\theta\}_{B^+_>}+i\eps+i\pi,\{\mu\}_{I_{p,<}},\{\theta\}_{A^-_<})
 F^{\ordo_2}(\{\mu\}_{I_{p,>}}+i\pi,\{\theta\}_{B^-_>}+i\eps+i\pi ,\{\theta\}_{A^+_<}) \right]. \\
\end{split}
\end{equation}
We stress that it is important to keep the $i\eps$ shift in the
kinematical factors $K_{x,t}$, because they can combine with the poles the form
factors to produce terms proportional to $x$ and
$t$. Examples for this are given in Sec. \ref{sec:comp}. On the other
hand, the shifts could be omitted from the phase factors, because
these factors
only depend on the rapidity differences within a given set. It is also important that all
$\eps$ variables have to be identical, because any other choice with
some $\eps_j/\eps_k\ne 1$ would lead to different finite terms. 

\begin{figure}
  \centering
  \begin{tikzpicture}
\newcommand{\operator}[2]{    \draw [ultra thick] (#1+0,#2+0) circle [radius=0.25];
    \draw [thick] (#1-0.17,#2-0.17) -- (#1+0.17,#2+0.17);
      \draw [thick] (#1-0.17,#2+0.17) -- (#1+0.17,#2-0.17);
    }

    \operator{0}{0}
     \operator{5}{0}
     \draw [<->,thick] (0.35,0) -- (4.65,0);
     \draw [<->,thick] (0.35,0.1) to [out=5,in=175] (4.65,0.1);
         \draw [<->,thick] (0.35,-0.1) to [out=-5,in=185]
         (4.65,-0.1);

 \draw [<-,thick] (0.35,-0.2) to [out=-13,in=193]   (8.65,-0.2);
 \draw [<-,thick] (0.35,-0.3) to [out=-14,in=194] (8.65,-0.3);

  \draw [<-,thick] (-4.35,0.2) to [out=13,in=167]   (4.65,0.2);
  \draw [<-,thick] (-4.35,0.3) to [out=14,in=166] (4.65,0.3);

  \draw [<-,thick] (5.35,0.07) to [out=5,in=175] (8.65,0.07);
  \draw [<-,thick] (5.35,-0.07) to [out=-5,in=185] (8.65,-0.07);

    \draw [<-,thick] (-4.35,0.07) to [out=5,in=175] (-0.35,0.07);
   \draw [<-,thick] (-4.35,-0.07) to [out=-5,in=185] (-0.35,-0.07);

   \node at (-0.5,-0.5) {$\ordo_1(0,0)$};
   \node at (5.5,0.5) {$\ordo_2(x,t)$};

   \node at (7.5,-1) {$\{\theta_{A^-}\}$};
   \node at (-2.5,1) {$\{\theta_{B^-}+i\eps\}$};
   \node at (7.5,0.5) {$\{\theta_{A^+}\}$};
   \node at (-2.5,-0.5) {$\{\theta_{B^+}+i\eps\}$};
   \node at (1.5,0.45) {$\{\mu\}$};

   \node at (9.5,0) {$\ket{\{\Omega\}}$};
   \node at (-5.2,0)  {$\bra{\{\Omega\}}$};
  \end{tikzpicture}
  \caption{Graphical interpretation of the LM
    series \eqref{LM2ptD}. The set $\{\theta\}$ represents the
    particles from the representative state $\ket{\Omega}$ that
    interact with the two-point function, whereas $\{\mu\}$ are
    intermediate particles between the two operators. Each physical
    amplitude is associated with additional phase factors, that depend
  on the partitions (ordering of the particles) and the contour for
  the $\mu$-integrals.}
  \label{fig:graphint2}
\end{figure}
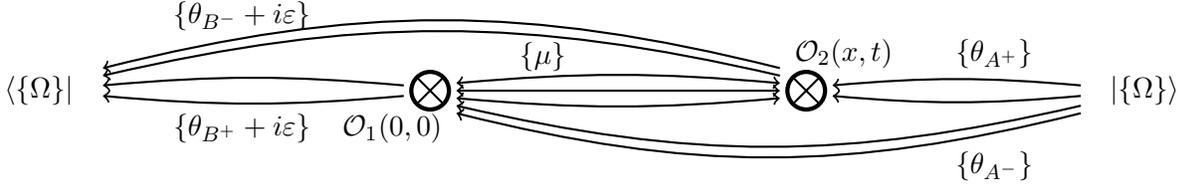

Representation \eqref{LM2ptD} is perhaps the most transparent from a physical
point of view: the rapidities $\{\theta\}$ stand for particles that
are present in the representative state $\ket{\Omega}$ and interact
with the two-point function, whereas the $\{\mu\}$ are additional
intermediate particles between the two operators. A graphical
interpretation is given in Figure \ref{fig:graphint2}.  
We also point out the striking
similarity with the original proposal
\eqref{eq:2point_leclair_mussardo}, however, there are crucial
differences. In \eqref{eq:2point_leclair_mussardo} the energy and
momenta of the intermediate particles is dressed, and it is not
specified how to deal with the kinematical poles in the higher order
terms. In contrast, here the energy and momenta are not dressed, and
all terms are regular due to the well-defined shifts $i\alpha$ and
$\eps$. The only ,,dressed'' quantities in \eqref{LM2ptD} are the
weight functions $f(\theta)$ and $\omega(\theta)$, and
the derivation in \cite{sajat-LM} shows that these are statistical weights
that are determined by the Bethe Ansatz description of the state
$\ket{\Omega}$.

To conclude this section we discuss the clustering property of the
two-point function. In the limit of large separations it is expected that
\begin{equation}
\lim_{|x|\to \infty}  \bra{\Omega}\ordo_1(0)\ordo_2(x,t)\ket{\Omega}=
\bra{\Omega}\ordo_1\ket{\Omega}   \bra{\Omega}\ordo_2\ket{\Omega}.
 \end{equation}
 This identity is motivated by physical requirements about the pure
 state $\ket{\Omega}$, but its explicit confirmation 
 is a highly non-trivial test of the various integral formulas
 \cite{D22,takacs-szecsenyi-2p}. Here we prove that the LM series
 satisfies the clustering property; the simplest way is to use the
 representation \eqref{LM2ptD}.

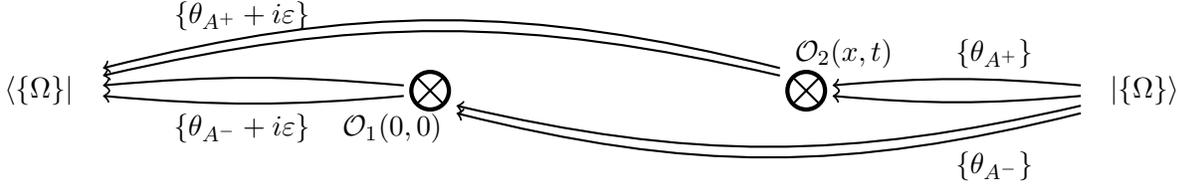
\begin{figure}
  \centering
  \begin{tikzpicture}
\newcommand{\operator}[2]{    \draw [ultra thick] (#1+0,#2+0) circle [radius=0.25];
    \draw [thick] (#1-0.17,#2-0.17) -- (#1+0.17,#2+0.17);
      \draw [thick] (#1-0.17,#2+0.17) -- (#1+0.17,#2-0.17);
    }

    \operator{0}{0}
     \operator{5}{0}
%     \draw [<->,thick] (0.35,0) -- (4.65,0);
 %    \draw [<->,thick] (0.35,0.1) to [out=5,in=175] (4.65,0.1);
  %       \draw [<->,thick] (0.35,-0.1) to [out=-5,in=185]
    %     (4.65,-0.1);

 \draw [<-,thick] (0.35,-0.2) to [out=-13,in=193]   (8.65,-0.2);
 \draw [<-,thick] (0.35,-0.3) to [out=-14,in=194] (8.65,-0.3);

  \draw [<-,thick] (-4.35,0.2) to [out=13,in=167]   (4.65,0.2);
  \draw [<-,thick] (-4.35,0.3) to [out=14,in=166] (4.65,0.3);

  \draw [<-,thick] (5.35,0.07) to [out=5,in=175] (8.65,0.07);
  \draw [<-,thick] (5.35,-0.07) to [out=-5,in=185] (8.65,-0.07);

    \draw [<-,thick] (-4.35,0.07) to [out=5,in=175] (-0.35,0.07);
   \draw [<-,thick] (-4.35,-0.07) to [out=-5,in=185] (-0.35,-0.07);

   \node at (-0.5,-0.5) {$\ordo_1(0,0)$};
   \node at (5.5,0.5) {$\ordo_2(x,t)$};

   \node at (7.5,-1) {$\{\theta_{A^-}\}$};
   \node at (-2.5,1) {$\{\theta_{A^+}+i\eps\}$};
   \node at (7.5,0.5) {$\{\theta_{A^+}\}$};
   \node at (-2.5,-0.5) {$\{\theta_{A^-}+i\eps\}$};

   \node at (9.5,0) {$\ket{\{\Omega\}}$};
   \node at (-5.2,0)  {$\bra{\{\Omega\}}$};
  \end{tikzpicture}
  \caption{Graphical interpretation of the clustering property of the
    two-point function. In the $|x|\to\infty$ limit only those terms
    survive in the two-point function, where $\{\mu\}=\emptyset$ and
    the partitionings of the index sets $A$ and $B$ are complementary,
    leading to a factorization of the two-point function.}
  \label{fig:graphint3}
\end{figure}

 The kinematical factors $K_{x,t}$ include multipliers of the form
 $e^{\pm ip(\theta)x}$, which result in terms of $\ordo(e^{-m|x|})$
 after integration in $\theta$. Therefore, in the large distance limit
 only those terms survive where all $K$-factors are trivial. This
 happens when there are no intermediate particles, and
  the  partitions have to satisfy $A^+=B^-$ and $A^-=B^+$. In these cases the
incoming and outgoing $\theta_j$ rapidities are always attached to the
same operator, and we obtain
\begin{equation}
  \label{LM2ptD2}
  \begin{split}
 & \lim_{|x|\to\infty}\bra{\Omega}\ordo_1(0)\ordo_2(x,t)\ket{\Omega}=\lim_{\eps\to 0}
 \left[\sum_{n=0}^\infty \frac{1}{n!}
\prod_{j=1}^n \int_\valos \frac{d\theta_j}{2\pi} f(\theta_j)\omega(\theta_j)
 \right.\times \\
 &\times \sum_{A^+\cup A^-=I_n}
  S^{<}(\{\theta\}_{A^+_<}|\{\theta\}_{A^-_<})
S^{>}(\{\theta\}_{A^-_>}|\{\theta\}_{A^+_>})\\
&\times \left.F^{\ordo_1}(\{\theta+i\eps\}_{A^-_>}+i\pi,\{\theta\}_{A^-_<})
 F^{\ordo_2}(\{\theta+i\eps\}_{A^+_>}+i\pi ,\{\theta\}_{A^+_<}) \right]. \\
\end{split}
\end{equation}
The phase factors cancel each other by definition, and then the integrals
completely factorize according to the partitions. 
By permutation symmetry we
obtain
\begin{equation}
  \label{LM2ptD3}
  \begin{split}
    & \lim_{|x|\to\infty}\bra{\Omega}\ordo_1(0)\ordo_2(x,t)\ket{\Omega}=\\
&    \lim_{\eps\to 0}
 \left[\sum_{n=0}^\infty \sum_{k=1}^n \frac{1}{k!(n-k)!}\right.
\prod_{j=1}^k \int_\valos \frac{d\theta_j}{2\pi} f(\theta_j)\omega(\theta_j)
 \times  F^{\ordo_1}(\{\theta+i\eps\}_{I_k,>}+i\pi,\{\theta\}_{I_k,<})\times \\
 &\times
 \left.
\prod_{j=1}^{n-k} \int_\valos \frac{d\tilde\theta_j}{2\pi} f(\tilde\theta_j)\omega(\tilde\theta_j)
   F^{\ordo_2}(\{\tilde\theta+i\eps\}_{I_{n-k,>}}+i\pi
   ,\{\tilde\theta\}_{I_{n-k},<}) \right]=
 \bra{\Omega}\ordo_1\ket{\Omega}   \bra{\Omega}\ordo_2\ket{\Omega}.
\end{split}
\end{equation}
A graphical interpretation of this identity is shown in Fig. \ref{fig:graphint3}.

%----------------------------------------------------------------------------------------------------------------------
%%%%%%%%%%%%%%%%%%%%%%%%%%%%%%%%%%%%%%%%%%%%%
%----------------------------------------------------------------------------------------------------------------------

\section{Comparison to our earlier work}

\label{sec:comp}

Here we compare the formula \eqref{LMmaskepp} to our previous results
for the finite temperature case \cite{D22,takacs-szecsenyi-2p}. These
articles approached the evaluation of finite temperature two-point
functions through finite volume regularization, leading to a double
series
\begin{equation}
  \label{regi}
\bra{\Omega}\ordo_1(0,0)\ordo_2(x,t)\ket{\Omega}=\sum_{N,M}D_{N,M}.
\end{equation}
Here $\ket{\Omega}$ is a representative state of the Gibbs
ensemble, and the r.h.s. is the result of a linked cluster expansion,
where the $D_{N,M}$ are well defined $L\to\infty$ limits of finite volume regularized
expressions.
For the details of the calculation we refer the reader to the
original articles; here we only describe the main properties of the
expansion \eqref{regi} and cite a few relevant formulas.

The series \eqref{regi} is organized as a low temperature expansion, such that each
$D_{N,M}$ carries a thermal weight $e^{-NmR \cosh(\theta)}$ and $M$ refers to the number of
intermediate particles in the finite volume regularization scheme. 
It is
important that $N$ does not correspond to the index $n$ in
\eqref{LM2pt}: the LeClair-Mussardo series involves the
weights $f(\theta)=1/(1+e^{\eps(\theta)})$, which have a
low-temperature expansion themselves. Therefore the terms with a given
$n$ in \eqref{LM2pt} include certain terms from $D_{N,M}$ with
$N>n$. Similarly, although the index $M$
plays a similar role as the number 
$p$ of intermediate integrals in
\eqref{eq:G_nm_final_contour+}-\eqref{eq:G_nm_final_contour-}, the
terms can not be matched one to one. This is an effect of the differences in
the  regularization
schemes, and it can be seen explicitly in the example presented below.

We compare the LM series to \eqref{regi}
 up to first
order in $e^{-mR \cosh(\theta)}$. Due to the infinite number of
possible intermediate particles between the two operators this is
already a strong independent confirmation of \eqref{LM2pt}.

In the
first order approximation the \eqref{TBA} gives simply 
$f(\theta)=e^{-mR \cosh(\theta)}$ and we expect to match the zeroth
and first order contributions as
\begin{equation}
\begin{split}
\sum_{M}D_{0M}=G^{x,t}_{0,c}(\emptyset)
\end{split}
\end{equation}
\begin{equation}
  \label{D1Mcheck}
\sum_{M}D_{1M}=\int \frac{\mathrm{d}\theta}{2\pi}e^{-mR\cosh(\theta)}
 G^{x,t}_{1,c}(\theta).
\end{equation}
These two terms are evaluated in the following two subsections. 

\subsection{$\GG_{0}^c$}

We have simply
\begin{equation}
  G^{x,t}_{0,c}(\emptyset)=
  \sum_{p=0}^\infty\frac{1}{p!}\left(\prod_{j=1}^p
    \int_{\valos+i\alpha} \frac{\mathrm{d}\mu_j}{2\pi}\right)
  K_{x,t}(\{\mu\}|\emptyset) F^{\ordo_1}\big(\{\mu\}_<\big)F^{\ordo_2}\big(\{\mu\}_>+i\pi\big)\,,
\end{equation}
which is nothing more that the spectral series for the
zero-temperature two-point function. In this case there is no
singularity for the $\mu$-variables, and the integration contour could
be pulled back to the real line. The same result was given for $\sum_M
D_{0M}$ in \cite{D22}.

\subsection{$\GG_{1}^c$}

We calculate the connected part of $\GG_{1}$,  which is defined as the
$\eps$ independent part of the bi-local form factor $\GG_{1}(\theta+i
\eps|\theta)$. Two-particle form factors don't have kinematical poles,
therefore in this case the connected part coincides simply with the
$\eps\to 0$ limit. From \eqref{eq:G_nm_final_contour+} we have the integral representation
\begin{equation}
  \label{GG1c}
\begin{split}
\GG_{1}(\theta+i \eps|\theta)=&\sum_{p=0}^\infty\frac{1}{p!}\left(\prod_{j=1}^p
  \int_{\valos+i\alpha}\frac{\mathrm{d}u_j}{2\pi}\right) \\
 \times&\Big\{K_{x,t}(\{u\}|\theta) F^{\ordo_1}\big(\theta+i\pi+i\eps,\{u\}_<\big)F^{\ordo_2}\big(\{u\}_>+i\pi,\theta \big) \\
 &+K_{x,t}(\{u\}|\emptyset) F^{\ordo_1}\big(\theta+i\pi+i\eps,\{u\}_<,\theta\big)F^{\ordo_2}\big(\{u\}_>+i\pi\big) \\
 &+K_{x,t}(\{u\},\theta+i\eps|\theta) F^{\ordo_1}\big(\{u\}_<\big)F^{\ordo_2}\big(\{u\}_>+i\pi,\theta+i\pi+i\eps,\theta\big) \\
 &+K_{x,t}(\{u\},\theta+i\eps|\emptyset) F^{\ordo_1}\big(\{u\}_<,\theta \big)F^{\ordo_2}\big(\{u\}_>+i\pi,\theta +i\pi +i\eps\big) \Big\}\,.
\end{split}
\end{equation}
The first and fourth terms are finite in the $\eps\to 0$ limit, but
the second and third terms have apparent
first order poles at $\eps=0$. These singularities cancel each other,
and care needs to be taken to obtain the finite left-over pieces.

We introduce an expansion of the form factors to make the singularities apparent
\begin{equation}
\begin{split}
F^{\ordo_1}\big(\theta+i\pi,\{u\}_{I,<}\big)=&F^{\ordo_1}_{reg}\big(\theta+i\pi,\{u\}_{I,<}\big)\\
+&i\sum_k\frac{\prod_{j<k}S(u_j-u_k)\left[1-\prod_{j\neq
      k}S(u_k-u_j)\right]
F^{\ordo_1}\big(\{u\}_{I\setminus\{k\},<}\big)}{\theta-u_k}\,, \\
F^{\ordo_2}\big(\{u\}_{I,>}+i\pi,\theta\big)=&F^{\ordo_2}_{reg}\big(\{u\}_{I,>}+i\pi,\theta\big)\\
+&i\sum_k\frac{\prod_{j>k}S(u_j-u_k)\left[1-\prod_{j\neq
      k}S(\theta-u_j)\right]
  F^{\ordo_2}\big(\{u\}_{I\setminus\{k\},>}+i\pi\big)}{\theta-u_k}\,. \\
\end{split}
\end{equation}
With the help of the expansion we can express the $\eps$ independent
part 
of the integrands in the second and third lines of \eqref{G1c}. Following \cite{D22,takacs-szecsenyi-2p} these
terms will also be called the ``connected parts''. They are given by
\begin{equation}
\begin{split}
&  F^{\ordo_1}_c\big(\theta+i\pi,\{u\}_{I,<},\theta\big)\equiv
  \mathrm{F.P.} \left(
    F^{\ordo_1}_c\big(\theta+i\pi+i\eps,\{u\}_{I,<},\theta\big)\right)
=F^{\ordo_1}_{reg}\big(\theta+i\pi,\{u\}_{I,<},\theta\big)+\\
  &+i\sum_k F^{\ordo_1}\big(\{u\}_{I\setminus\{k\},<},\theta\big)
  \frac{\prod_{j<k}S(u_j-u_k)\left[1-\prod_{j\neq k}S(u_k-u_j)S(u_k-\theta)\right]}{\theta-u_k}\,, \\
\end{split}
\end{equation}

\begin{equation}
  \label{hujj}
  \begin{split}
&\tilde{F}^{\ordo_2}_c\big(\{u\}_{I,>}+i\pi,\theta+i\pi|\theta\big)\equiv\frac{\mathrm{F.P.}\left( K_{x,t}(\{u\}_I,\theta+i\eps|\theta) F^{\ordo_2}\big(\{u\}_{I,>}+i\pi,\theta+i\pi+i\eps,\theta\big)\right)}{K_{x,t}(\{u\}_I |\emptyset)} \\
&=F^{\ordo_2}_{reg}\big(\{u\}_{I,>}+i\pi,\theta+i\pi,\theta\big)+
\\
&+i\sum_k F^{\ordo_2}\big(\{u\}_{I\setminus\{k\},>}+i\pi,\theta+i\pi \big)
\frac{\prod_{j>k}S(u_j-u_k)\left[1+\prod_{j\neq k}S(\theta-u_j)\right]}{\theta-u_k}+ \\
&+\left[1-\prod_{j\in I}S(u_j-\theta)\right]\left(\sum_{j \in
    I}\fii(u_j-\theta)+i[t \partial E(\theta)+ix\partial P(\theta)]\right)  F^{\ordo_2}\big(\{u\}_{I\setminus\{k\},>}+i\pi\big)\,.
\end{split}
\end{equation}
In the \eqref{hujj} the tilde denotes, that the connected form factor
incorporates the contribution form the energy and momentum phase
factor as well.  

With the regular pieces introduced above the connected bi-local form
factor is written as
\begin{equation}
  \label{G1c}
\begin{split}
\GG_{1}^{c}(\theta|\theta)=& \sum_{a=0}^\infty\frac{1}{a!}\left(\prod_{j=1}^a  \int_{\valos+i\alpha}\frac{\mathrm{d}u_j}{2\pi}\right) \\
 \times&\Big\{K_{x,t}(\{u\}|\theta) F^{\ordo_1}\big(\theta+i\pi,\{u\}_<\big)F^{\ordo_2}\big(\{u\}_>+i\pi,\theta \big) \\
 &+K_{x,t}(\{u\}|\emptyset) F^{\ordo_1}_c\big(\theta+i\pi |\{u\}_<,\theta\big)F^{\ordo_2}\big(\{u\}_>+i\pi\big) \\
 &+K_{x,t}(\{u\}|\emptyset) F^{\ordo_1}\big(\{u\}_<\big)\tilde{F}^{\ordo_2}_c\big(\{u\}_>+i\pi,\theta+i\pi ,\theta\big) \\
 &+K_{x,t}(\{u\},\theta|\emptyset) F^{\ordo_1}\big(\{u\}_<,\theta \big)F^{\ordo_2}\big(\{u\}_>+i\pi,\theta +i\pi \big) \Big\}\,.
\end{split}
\end{equation}

The $D_{1M}$ contributions to \eqref{regi}  were calculated in  \cite{D22,takacs-szecsenyi-2p} as
\begin{equation}
\begin{split}
D_{10}=& \int_{\valos}\frac{\mathrm{d}\theta}{2\pi}e^{-mR\cosh(\theta)}K_{x,t}\left(\emptyset | \theta \right)F^{1}\left(\theta+i\pi\right)F^{2}\left(\theta\right)\,,\\
D_{11}=&\int_{\valos}\frac{\mathrm{d}\theta}{2\pi}e^{-mR\cosh(\theta)} \Bigg\{ F^{1}\left(\theta+i\pi,\theta\right)\left\langle \ordo_{2}\right\rangle +\left\langle \ordo_{1}\right\rangle F^{2}\left(\theta+i\pi,\theta\right) \\
&+\int_{\valos+i\alpha}\frac{\mathrm{d}u_{1}}{2\pi} K_{x,t}\left(u_{1} |\theta \right)F^{1}\left(\theta+i\pi,u_{1}\right)F^{2}\left(u_{1}+i\pi,\theta\right) \Bigg\}\,, \\
D_{1,M\ge2}=& \int_{\valos}\frac{\mathrm{d}\theta}{2\pi}e^{-mR\cosh(\theta)} \Bigg\{\frac{1}{M!}\left(\prod_{j=1}^M  \int_{\valos+i\alpha}\frac{\mathrm{d}u_j}{2\pi}\right)  K_{x,t}(\{u\}|\theta) F^{\ordo_1}\big(\theta+i\pi,\{u\}_<\big)F^{\ordo_2}\big(\{u\}_>+i\pi,\theta \big) \\
&+\frac{1}{(M-1)!}\left(\prod_{j=1}^{M-1}  \int_{\valos+i\alpha}\frac{\mathrm{d}u_j}{2\pi}\right)  K_{x,t}(\{u\}|\emptyset) F^{\ordo_1}_c\big(\theta+i\pi|\{u\}_<,\theta\big)F^{\ordo_2}\big(\{u\}_>+i\pi \big) \\
&+\frac{1}{(M-1)!}\left(\prod_{j=1}^{M-1}  \int_{\valos+i\alpha}\frac{\mathrm{d}u_j}{2\pi}\right)  K_{x,t}(\{u\}|\emptyset) F^{\ordo_1}\big(\{u\}_<,\big)\tilde{F}^{\ordo_2}_c\big(\{u\}_>+i\pi,\theta+i\pi,\theta \big) \\
&+ \frac{1}{(M-2)!}\left(\prod_{j=1}^{M-2}  \int_{\valos+i\alpha}\frac{\mathrm{d}u_j}{2\pi}\right)  K_{x,t}(\{u\},\theta|\emptyset) F^{\ordo_1}\big(\{u\}_<,\theta\big)F^{\ordo_2}\big(\{u\}_>+i\pi,\theta+i\pi \big)\Bigg\}\,.
\end{split}
\end{equation}
Comparing the form of $D_{1M}$ to $\GG_1^{c}$ we see, that 
\begin{equation}
\sum_{M}D_{1M}=\int \frac{\mathrm{d}\theta }{2\pi}e^{-mR\cosh(\theta)}\GG_{1}^c(\theta|\theta),
\end{equation}
as required. Note that $D_{1,M}$ includes terms that have $M$, $M-1$ and
$M-2$ intermediate integrals, whereas in \eqref{G1c} the number of
integrals for the terms with a
given $p$ is always $p$. Nevertheless, the summed quantities exactly
correspond to each other.

\section{Conclusions}

\label{sec:conclusions}

In this work we derived a LeClair-Mussardo formula for the two-point
function at space-like separations. The general structure of the series is given by the two implicit
representations 
\eqref{LM2pt} and \eqref{LM2ptB}.
These formulas involve the form factors of the
bi-local operators, for which the integral series are given by
\eqref{eq:G_nm_final_contour+} or \eqref{eq:G_nm_final_contour-},
depending on the sign of $x$.
More explicit representations are given by \eqref{LMmaskepp}, and
finally \eqref{LM2ptD}. We believe that the latter form is the
physically most transparent; its graphical interpretation is plotted
on Fig. \ref{fig:graphint2}. The result can be considered as a
large distance expansion, because each intermediate particle (each
$\mu$-integral) carries a weight of $\sim e^{-ms}$ with
$s^2=x^2-t^2$. 

It is an important property of the final formulas that each term is explicit, with a well
defined prescription to deal with the kinematical singularities. This
was missing in the previous works: the original proposal of
\cite{leclair_mussardo} had singularities in the higher order terms,
and the regularization of \cite{D22,takacs-szecsenyi-2p} was only
performed on a case by case basis.

Our results are expressed in terms of the bare form factors of the 
theory, and the momenta and energy of the intermediate particles between
the operators are also the bare quantities. This follows from our
approach to treat the bi-local operator as a composite object, and to
develop the LM series based on the bi-local form factors. 
The only dressed quantities are the statistical filling fractions
$f(\theta)$ and the weight function $\omega(\theta)$. These
two are derived from
the Bethe Ansatz description of the background $\ket{\Omega}$, and
they are the same for all local and bi-local operators. 

The original proposal of  \cite{leclair_mussardo} used the bare form
factors of the theory, but it suggested a dressing for the
intermediate particles. This might seem intuitive, but our
derivation shows that it is inconsistent. We 
expressed the two point function with the bare quantities, and if
there is a partial re-summation of the series, then it should lead to a
dressing for both the form factors and the kinematical
factors. Examples for such dressing can be found in the context of
non-relativistic models \cite{Korepin-book}, and a similar structure
was also implied by the results of \cite{Doyon-GHD-LM}, at least in
the large time limit.

The methods and results of this paper also apply to non-relativistic models, such as the
1D Bose gas. In fact, most of these results were already derived
for the Bose gas in
\cite{Thacker-elso,PhysRevD.21.1523,creamer-thacker-wilkinson-81,creamer-thacker-wilkinson-rossz,PhysRevD.23.3081},
and these papers served as a motivation for the present work. The main
addition of our work is the realization that the kinematical pole
structure of the bi-local operator is the same even in the
relativistic case, and that this property is enough to establish the
LM series. We presented three arguments for the kinematical pole;
the most rigorous is the explicit check presented in \ref{sec:kin}. 

The form factors of the bi-local operator were obtained by inserting a
complete set of states between the two operators. This is the same
technique that was used by Smirnov \cite{smirnov_ff}, which lead to
the proof of the local commutativity theorem. There is one essential
step in both calculations: a certain shift of integration contours,
which is possible only for space-like separations.

As tests of our results for the LM series we evaluated the first order corrections in the
low-temperature expansion in Section \ref{sec:comp} and compared them to
existing results; complete agreement was found for an arbitrary number of
intermediate particles. As an additional test we performed the large
distance limit, and confirmed that the integral series factorizes into
the product of two LeClair-Mussardo series, thus fulfilling the
clustering property.  This is a highly nontrivial test of the final
formula \eqref{LM2ptD}.

The most important open problem is to find methods for 
 the actual implementation of
the final formulas. As shown by the examples in Sec. \ref{sec:comp},
this can be a cumbersome task, both numerically and
analytically.

It is also important to consider the case of time-like
separations. Our first two arguments for the validity kinematical pole
property (those based on the OPE and the Bethe wave function) do
not apply in this situation, but preliminary calculations show that
the integral series for the bi-local form factor can be modified such
that the kinematical pole property can be proven
nevertheless. However, the convergence properties of the 
resulting integral series can be drastically different, and we don't
have a decisive answer whether a LM series can be established for this
case too.

Coming back to space-like separations, it would be interesting to consider the first corrections in a large
distance expansion. Generally it is expected that
\begin{equation}
  \label{longd}
  \bra{\Omega}\ordo_1(0)\ordo_2(x,0)\ket{\Omega}=
  \bra{\Omega}\ordo_1\ket{\Omega}  \bra{\Omega}\ordo_2\ket{\Omega}+ \ordo(e^{-x/\xi}),
\end{equation}
where $\xi$ is a correlation length.
In the thermal situation Euclidean invariance implies, that the
 above object is equal to a two-point function with time-like
separation, evaluated in the ground state of a finite volume QFT with
volume $R=1/T$.
In this case the correlation length is $1/\xi=E_1-E_0$,
where $E_0$ and $E_1$ are the exact ground state and first excited state
energies in the given volume. It is an
interesting question whether these quantities (or possibly even the exact
finite volume transition matrix elements that determine the pre-factors of the
correction terms in \eqref{longd}) could be extracted from the LM
series.

Finally, it would be interesting to establish a connection to the results
of \cite{Doyon-GHD-LM}. Equation (3.38) in that work is reminiscent of our
formulas, even though it was derived for time-like separations in the
large time limit. The concrete relations between the two integral
series have yet to be investigated.

We hope to return to these questions in future research.

\bigskip

Note added: After this work was finished we became aware of the work
\cite{zoli-ff-luscher} which derives the so-called first Lüscher's
corrections to finite volume form factors in iQFT. The methods and
results there are related to our work. In
\cite{zoli-ff-luscher} the new results were shown to be consistent with the previous
works  \cite{D22,takacs-szecsenyi-2p}, thus they are consistent with our present
results too.

\vspace{1cm}
{\bf Acknowledgments} 

We would like to thank G\'abor Tak\'acs, Benjamin Doyon, Zolt\'an
Bajnok, and Lorenzo Piroli for useful discussions. 

B.P. acknowledges support from the ``Premium'' Postdoctoral
Program of the Hungarian Academy of Sciences, the K2016 grant
no. 119204 and the KH-17 grant no. 125567 of the research agency NKFIH. 
This work was partially supported also within the Quantum
Technology National Excellence Program (Project No. 2017-1.2.1-NKP-2017-00001).

I.M.SZ. work was partially supported by the EPSRC Standard Grant "Entanglement Measures, Twist Fields, and Partition Functions in Quantum Field Theory" under reference number EP/P006108/1, and has received funding from the People Programme (Marie Curie Actions) of the European Union’s Seventh Framework Programme FP7/2007- 2013/ under REA Grant Agreement No 317089 (GATIS).

\bigskip

\appendix

\section{Contour transformation}

\label{appA}

Here we perform the contour transformation in the $\mu$-variables in the
original formula \eqref{eq:G_nm_final_original_contour}. The goal is
to shift the contour to $\valos+i\alpha$, thereby hitting the
kinematical singularities whenever the $\mu$-integrals cross one of
the $\vartheta$ variables. 
 
Let us focus on a term in $G^{x,t}_{n,m}$ with given $A^\pm$ and $B^\pm$ subsets, and the residue, when $r\le |B^+|$ number of $\mu$s coincide with same number of $\vartheta$s, denoted by the set $\tilde{B}$. Due to the exchange axiom \eqref{eq:exchangeaxiom} and relabeling the integration variables, we only calculate the residue when $\vartheta_{\tilde{b}_i}=\mu_i$ and take into account the other permutation by the combinatorial factor $p!/(p-r)!$. Furthermore, we use the notation $\hat{B}=B^+\setminus\tilde{B}$ and $\tilde{\mu}_i=\mu_{i+r}$. The residue to evaluate is
\begin{equation}
\begin{split}
  &\sum_{p=r}^\infty \frac{1}{(p-r)!}\int_{\valos+i\alpha}
  \prod_{i=1}^{p-r}\frac{\mathrm{d}\tilde{\mu}_{i}}{2\pi}
  \sum_{\tilde{B}\cup\hat{B}=B^+} \prod_{i=1}^r \int_{\mathcal{C}_i}\frac{\mathrm{d}\mu_{i}}{2\pi}   
 K_{x,t}(\{\mu\}_{I_r},\{\tilde{\mu}\}_{I_{p-r}},\{\vartheta\}_{B^-}|\{\theta\}_{A^+})\\
&\times F^{\mathcal{O}_1}(\{\vartheta\}_{B^+_>}+i\pi ,\{\mu\}_{I_{r,<}},\{\tilde{\mu}\}_{I_{p-r,<}},\{\theta\}_{A^-_<})\\
&\times F^{\mathcal{O}_2}(\{\vartheta\}_{B^-_>}+i\pi ,\{\tilde{\mu}\}_{I_{p-r,>}}+i\pi,\{\mu\}_{I_{r,>}}+i\pi,\{\theta\}_{A^+_<})  \\
&\times S^{<}(\{\theta\}_{A^+_<}|\{\theta\}_{A^-_<}) S^{>}(\{\vartheta\}_{B^-_>}|\{\vartheta\}_{B^+_>})  S^\leftrightarrow(\{\vartheta\}_{B^-_>}|\{\theta\}_{A^-_<})\,,
\end{split}
\end{equation}
where $\mathcal{C}_i$ is contour encircling $\vartheta_{\tilde{b}_i}$ with positive orientation. We omitted the $\pm i0$ shifts in the formula, since there are no more poles on the integration contours, the form factor is off-diagonal ($\vartheta_i\neq\theta_j$) and the shifts are not necessary anymore. We note, that in the case $|A^-|=0$, $p=|B^+|$ and $p=r$, the residue is missing, since the two-particle form factor is regular. With \eqref{eq:exchangeaxiom} and \eqref{eq:kinematicalaxiom} the residue evaluates to
\begin{equation}
\begin{split}
&\sum_{p=r}^\infty \frac{1}{(p-r)!}\int_{\valos+i\alpha}\prod_{i=1}^{p-r}\frac{\mathrm{d}\tilde{\mu}_{i}}{2\pi} \sum_{\tilde{B}\cup\hat{B}=B^+}   K_{x,t}(\{\vartheta\}_{\tilde{B}},\{\tilde{\mu}\}_{I_{p-r}},\{\vartheta\}_{B^-}|\{\theta\}_{A^+})\\
&\times F^{\mathcal{O}_1}(\{\vartheta\}_{\hat{B}_>}+i\pi, \{\tilde{\mu}\}_{I_{p-r,<}},\{\theta\}_{A^-_<})\\
&\times S^{>}(\{\vartheta\}_{\hat{B}_>}|\{\vartheta\}_{\tilde{B}_>}) \prod_{i=1}^r[1-S^\leftrightarrow(\{\vartheta\}_{\tilde{b}_i}|\{\vartheta\}_{\hat{B}_<}+i\pi)S^\leftrightarrow(\{\vartheta\}_{\tilde{b}_i}|\{\tilde{\mu}\}_{I_{p-r,<}})S^\leftrightarrow(\{\vartheta\}_{\tilde{b}_i}|\{\theta\}_{A^-_<})]
\\
&\times F^{\mathcal{O}_2}(\{\vartheta\}_{B^-_>}+i\pi ,\{\tilde{\mu}\}_{I_{p-r,>}}+i\pi,\{\vartheta\}_{\tilde{B}_>}+i\pi,\{\theta\}_{A^+_<})  \\
&\times S^{<}(\{\theta\}_{A^+_<}|\{\theta\}_{A^-_<}) S^{>}(\{\vartheta\}_{B^-_>}|\{\vartheta\}_{B^+_>})  S^\leftrightarrow(\{\vartheta\}_{B^-_>}|\{\theta\}_{A^-_<})\,.
\end{split}
\end{equation}
The next step is to expand the product in the formula. We separate the
set $\tilde{B}$ into two disjoint subsets
$\tilde{B}_I\cup\tilde{B}_S=\tilde{B}$. For indices in $\tilde{B}_I$
we pick the $1$-term in the product, for the indices in $\tilde{B}_S$
we pick the product of the S-matrices. Using the exchange axiom
\eqref{eq:exchangeaxiom}
we simplify the expression to 
\begin{equation}
\label{eq:residue_from_shift_gen}
\begin{split}
&\sum_{p=0}^\infty \frac{1}{p!}\int_{\valos+i\alpha}\prod_{i=1}^{p}\frac{\mathrm{d}\mu_{i}}{2\pi} \sum_{\tilde{B}\cup\hat{B}=B^+} \sum_{\tilde{B}_I\cup\tilde{B}_S=\tilde{B}} (-1)^{|\tilde{B}_S|}\\
&\times F^{\mathcal{O}_1}(\{\vartheta\}_{\hat{B}_>}+i\pi, \{\mu\}_{I_{p,<}},\{\theta\}_{A^-_<})   
K_{x,t}(\{\vartheta\}_{\tilde{B}_I},\{\vartheta\}_{(B^-\cup\tilde{B}_S)},\{\mu\}_{I_{p}}|\{\theta\}_{A^+})\\
&\times F^{\mathcal{O}_2}(\{\vartheta\}_{(B^-\cup\tilde{B}_S)_>}+i\pi,\{\mu\}_{I_{p,>}}+i\pi,\{\vartheta\}_{\tilde{B}_{I,>}}+i\pi,\{\theta\}_{A^+_<}) \\
&\times 
S^\leftrightarrow(\{\vartheta\}_{(B^-\cup\tilde{B}_S)_>}|\{\theta\}_{A^-_<})
S^{>}(\{\vartheta\}_{(B^-\cup\tilde{B}_S)_>}|\{\vartheta\}_{\hat{B}_>}) 
S^{>}(\{\vartheta\}_{\tilde{B}_{S,>}}|\{\vartheta\}_{\tilde{B}_{I,>}})
 \\
&\times S^{>}(\{\vartheta\}_{B^-_>}|\{\vartheta\}_{\tilde{B}_{I,>}}) S^{<}(\{\theta\}_{A^+_<}|\{\theta\}_{A^-_<}) 
S^{>}(\{\vartheta\}_{\hat{B}_>}|\{\vartheta\}_{\tilde{B}_{I,>}})\,,
\end{split}
\end{equation}
where for the set $B^-\cup\tilde{B}_S$ we imply the rearrangement of the elements to increasing order, and we relabeled the integration variables. 

It is important to note that apart from the sign of the expression, it
only depends on the sets $\hat{B}$, $\tilde{B}_I$ and
$B^-\cup\tilde{B}_S$ and $p$.  Since we have to sum for all disjoint
$B^+$ and $B^-$ that sums up to $B$, there is a chance for
cancellation of terms. For fixed $\hat{B}$, $\tilde{B}_I$ and
$B^-\cup\tilde{B}_S$ and $p$ the sum of the sign pre-factors is 
\begin{equation}
\sum_{|\tilde{B}_S|=0}^{|B^-\cup \tilde{B}_S|} \binom{|B^-\cup \tilde{B}_S|}{|\tilde{B}_S|} (-1)^{|\tilde{B}_S|} =\delta_{|B^-\cup \tilde{B}_S|,0}\,.
\end{equation}
After this cancellation we relabel the sets as $\hat{B}\to B^+$ and
$\tilde{B}_I\to B^-$ to arrive to the result \eqref{eq:G_nm_final_contour+}.

We should also investigate the situation when $|A^-|=0$, $p=|B^+|$ and
$p=r$. The general formula for the residues
\eqref{eq:residue_from_shift_gen} is not valid in this case, since the
two-particle form factor is regular, and we don't get corrections from
residues to the $|\hat{B}|=0$, $|A^-|=0$ and $p=0$ term. However,
these terms already are in the shape of
\eqref{eq:G_nm_final_contour+}, hence the final formula is valid.

\addcontentsline{toc}{section}{References}
\providecommand{\href}[2]{#2}\begingroup\raggedright\endgroup

\end{document}